\documentclass[prb,aps,twocolumn,showpacs]{revtex4}

\usepackage{graphics}
\usepackage{graphicx}
\usepackage{amssymb,amsmath}
\usepackage{bm}

\newcommand{\beq}{\begin{equation}}
\newcommand{\eeq}{\end{equation}}
\newcommand{\bea}{\begin{eqnarray}}
\newcommand{\eea}{\end{eqnarray}}

\begin{document}

\title{Kane-Mele Hubbard model on a zigzag ribbon: 
stability of the topological edge states and quantum phase transitions}
\author{ Chung-Hou Chung$^{1,2}$, Der-Hau Lee$^{1}$ and Sung-Po Chao$^{2,3}$}
\affiliation{
$^{1}$Electrophysics Department, National Chiao-Tung University,
HsinChu, Taiwan, 300, R.O.C. \\
$^{2}$Physics Division, National Center for Theoretical Sciences, HsinChu, Taiwan, 300 R.O.C.\\
$^{3}$Department of Physics, National Tsing-Hua University, HsinChu, Taiwan, 300 R.O.C.
 }
\date{\today}

\begin{abstract}
We study the quantum phases and phase transitions of the Kane-Mele Hubbard 
(KMH) model on a zigzag ribbon of honeycomb lattice at a finite size via the 
weak-coupling renormalization group (RG) 
approach. In the non-interacting limit, the KM model is known to 
support topological edge states where electrons show helical property 
with orientations of the spin and momentum being locked. 
The effective inter-edge hopping terms are generated due to finite-size effect. 
In the presence of an on-site Coulomb repulsive interaction and the 
inter-edge hoppings,  
special focus is put on the stability of the topological edge states 
(TI phase) in 
the KMH model against (i) the charge and spin gaped (II) phase, (ii) 
the charge gaped but spin gapless (IC) phase and (iii) 
the spin gaped but charge gapless (CI) phase depending on 
the number (even/odd) of the zigzag ribbons, 
doping level (electron filling factor) 
and the ratio of the Coulomb interaction to the inter-edge tunneling. 
We discuss different 
phase diagrams for even and odd numbers of zigzag ribbons. 
We find the TI-CI, II-IC, and II-CI  
quantum phase transitions are of the Kosterlitz-Thouless (KT) 
type.  
By computing various correlation functions, we further analyze 
the nature and leading instabilities of these phases.
\end{abstract}

\maketitle

%%%%%%%%%%%%%%%%%%%%%%%%%%%%%%%%%%%%%%%%%%%%%%%%%%%%%%%%%%%%%%%%%%%%%%%
\section{ Introduction.}

Recently, there has been growing interest in topological 
insulators (TIs) and superconductors which support gapless 
edge (surface) states while the bulk remains insulating
\cite{Kane-review,SCZhang-review}. 
These surface states come as a consequence of 
the spin-orbit (SO) couplings, and are protected by the 
time-reversal symmetry (TRS)\cite{Kane-review,SCZhang-review}. 
The topological nature of TIs 
lies in the non-trivial topological $Z_2$ invariant\cite{KM} while 
it becomes trivial for an ordinary band insulator (BI). 
The theoretical predictions\cite{SCZhang-science2006,Fu-Kane-2007,chinese-SCZhang} of TIs have been soon 
observed experimentally in various insulators with strong SO 
couplings\cite{TI-exp}.  
In two-dimensional systems, these topological states have been predicted 
in the framework of the quantum spin Hall insulator 
(QSHI)\cite{Haldane,KM,KM2,bernevig1,Fu,bernevig2}, and 
have been realized experimentally soon after in $HgTe/CdTe$ quantum well 
structures\cite{SCZhang-science2006}. Unlike the integer quantum Hall 
where the chiral (one propagating mode of electrons with a single spin 
species) edge states are generated by 
an external magnetic field which breaks TRS, 
the TRS preserving QSHI systems lead to helical edge states 
in the absence of a magnetic field in which propagation direction 
at one edge is opposite for opposite spins\cite{bernevig2}. 
This one-dimensional 
helical edge state electrons are protected by TRS\cite{KM} and are 
free of spin-flip backscatterings\cite{SCZhang-review}. 
As a result, they lead to 
perfect transmission in charge transport along the 
edge\cite{SCZhang-physicstoday}. 

A simple theoretical 
model was first introduced by Haldane\cite{Haldane} and later proposed 
by Kane and 
Mele\cite{KM,KM2} (the KM model) to capture the helical edge states of 
QSHIs. The 
KM model was aimed to describe edge states in graphene. Though the SO 
coupling in graphene is expected to be too small to observe the edge states, 
the KM Model is regarded as a generic model for 2D TIs. The existence of the 
helical edge states in KM model has been well studied. Recently, more 
attention has been put on the stability, exotic quantum phases and phase 
transitions 
of the helical edge states and 
 possible exotic quantum phases in the 
correlated Kane-Mele Hubbard\cite{Meng,hur,assaad,sandler,GMZhang} 
model upon including 
the on-site Coulomb repulsions (the Hubbard $U$ term)  
in the KM model. In a pioneering work by 
Meng {\it et al.} in Ref. ~\onlinecite{Meng} via Quantum Monte Carlo and dynamical mean-field approaches, 
the helical edge states are stable up to a finite Hubbard interaction, and 
a gaped spin-liquid phase was predicted in the phase diagram 
of the KM Hubbard model at half filling 
for small to intermediate range of $U$. Moreover, the 1D Luttinger liquid 
physics with power-law correlations 
for the helical edge states 
has been studied numerically\cite{assaad} as well as analytically\cite{bena} 
in the framework of the KM Hubbard model. Meanwhile, the doping effect on the 
KM Hubbard model was addressed in Ref.~\onlinecite{fiete} 
where spin liquid phase was argued 
to become superconducting state. 

\begin{figure}[t]
\begin{center}
%\vspace{0.2cm}
\includegraphics[angle=0,width=7cm,clip]{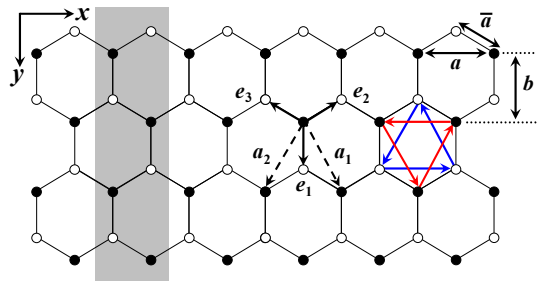}
\end{center}
\par
 \vskip -0.7cm
\caption{
(Color online)
Honeycomb lattice of a finite-sized zigzag ribbon of the 
tight-binding Kane-Mele model with the ribbon size $N=4$ ($N$ being the 
number of zigzag chains along $x-$axis) along $y-$axis. 
The honeycomb lattice consists of two inter-penetrating triangular 
lattices denoted by sublattice $A$ (dark circles) and sublattice 
$B$ (open circles) with lattice vectors ${\bf a_1}$ and ${\bf a_2}$ 
(dashed arrows).  
The zigzag ribbon shows translational symmetry along $x-$axis. 
The nearest-neighbor lattice vectors between nearest-neighbor $A$ and $B$ 
sites are denoted by ${\bf e_{i=1,2,3}}$ with a lattice constant $a$. The 
red (black) arrows within sublattice $A(B)$ represent the 
directions of the next-nearest-neighbor hopping term $\lambda_{SO}$ 
in the KM model (see text). The gray shaded region represents for the 
super-unit-cell of the zigzag ribbon, which repeats itself along $x-$axis.}
\label{ribbon}
\end{figure}

In this paper, we present a theoretical analysis on  
the KM Hubbard model at half filling and away from half filling  
from a different perspective: we analyze the 
model on a finite-sized zigzag ribbon 
(where the helical edge states 
have been realized numerically in the tight-binding KM model\cite{hur}) 
with a ribbon width $L=(N-1)b$ 
($N$ being the number of 
zigzag chain in a ribbon and $b$ is defined in Fig.~\ref{ribbon}) 
in the weak-coupling (weak on-site Coulomb $U$) 
limit via perturbative renormalization group (RG) combined with the 
bosonization approaches. Note that one can alternatively study 
the model on an armchair ribbon, which was suggested to support 
edge states in graphene 
(equivalent to the KM model without SO coupling)\cite{armchair}.
The authors in Ref.~\onlinecite{sandler} have studied 
the effects of long-range Coulomb interactions on the edge states of 
a finite-sized zigzag KM ribbon. The effects of the short-ranged 
electron-electron interaction on the helical 
edge states of the KM model have been addressed in Ref.~\onlinecite{Teo}. 
We shall emphasize here the stability of the helical 
edge states against the combined short-ranged on-site Coulomb 
interaction the and finite size effects, 
as well as possible other emerged quantum phases and phase transitions 
(QPTs)\cite{QPT} among them. 

The finite-size effect manifests itself in the structure of the 
energy spectrum and in an effective inter-edge tunneling terms. 
We further find that these behaviors for even number of zigzag KM ribbons 
($N=even$) are different from those for $N=odd$. 
For $N=even$, a finite energy gap is found at half filling where 
the Fermi energy is at the Dirac point 
$ka=\pi$. 
This small gap is due to breaking of the sublattice 
translational invariance at the boundaries,  
and can be explained in terms of 
an effective finite single-particle inter-edge tunneling, which 
decays exponentially with increasing $L$. 
Away form half filling, 
the energy dispersion becomes gapless at the Fermi level. 
For $N=odd$, however, the energy spectrum is gapless and the single-particle 
inter-edge tunneling vanishes for both 
half filling and away from half filling. Nevertheless, for both $N=even$ 
and $N=odd$, two-particle processes, effective 
inter-edge two-particle spin-flip and inter-edge 
Umklapp (two-particle backscattering) terms, are generated  
via second-order inter-edge hoppings. 

Our stability analysis of the KMH ribbon is summarized as follows. 
For $N=even$, the energy gap at half-filled at the Dirac point 
 gives rise to a charge and spin gaped (insulating) (II) phase\cite{Teo}; 
at a generic filling, however, the two-particle 
processes when combined with the 
effect of the Hubbard $U$ term lead to the instabilities of the helical 
edge states towards a charge gapless but spin gaped (CI) phase\cite{Teo} 
in the RG analysis via 
the Kosterlitz-Thouless type of quantum phase transitions. 
When $L\rightarrow \infty$, the inter-edge hopping term vanishes, 
the TI phase at half filling 
is unstable against the charge gaped but spin gapless 
(IC) phase\cite{Teo} for arbitrary $U>0$, while it is stable away 
from half filling.   
For $N=odd$, the single-particle inter-edge tunneling is absent, while 
the combined two-particle inter-edge hoppings and the on-site Coulomb 
repulsions make the TI unstable for any finite $U$ or inter-edge tunneling. 
As a result, the TI phase moves towards CI or IC or II  
phase depending on the ratio of Coulomb interaction and the 
inter-edge tunneling. The phase transitions for II-IC and II-CI are of the KT type.

By computing various 
correlation functions, we further 
analyze the instabilities of the helical edge states, the CI and IC phases 
towards the charge-density-wave (CDW), spin-density-wave 
(SDW) as well as the singlet (SS) and triplet (TT) superconducting states. 

The remaining parts of the paper is organized as follows. In Sec. II, 
the Kane-Mele Hubbard at a finite size is introduced. The model is 
re-expressed in terms of the scalar and vector current operators. 
In Sec. III, the stability of the helical edge states is addressed 
via weak-coupling RG analysis. We also address the nature of the 
quantum phase transitions between the TI and other quantum phases. 
We conclude in Sec. IV.

%%%%%%%%%%%%%%%%%%%%%%%%%%%%%%%%%%%%%%%%%%%%%%%%%%%%%%%%%%%%%%%%%%%%%%%
\section{ Model Hamiltonian.}

\subsection{The non-interacting Kane-Mele zigzag ribbon}

Before we study the interacting Kane-Mele Hubbard model, it is worthwhile 
summarizing the main results for the non-interacting Kane-Mele (KM) model 
on a zigzag ribbon of honeycomb lattice, 
given by the following Hamiltonian\cite{KM}:
\begin{eqnarray}
H_{KM} &=& -t\sum\limits_{\left\langle {ij} \right\rangle, \sigma } 
{c_{i\sigma}^\dagger c_{j\sigma}}   +
i\lambda _{SO} \sum\limits_{\left\langle {\left\langle {ij}
\right\rangle} \right\rangle,\sigma } {\nu _{ij} c_{i\sigma}^\dagger s^z
c_{j\sigma} } +  h.c.\nonumber \\
\label{H_KM}
\end{eqnarray} 
where $\left\langle i,j \right\rangle$ and $\left\langle \left\langle i,j \right\rangle\right\rangle$ refer to the nearest-neighbor (NN) and next-nearest-neighbor (NNN) sites, respectively. 
The NN and NNN 
lattice vectors for the honeycomb lattice are denoted respectively by 
${\bf e_{i=1,2,3}}$ and ${\bf a_{i=1,2}}$\cite{hur}:
\begin{eqnarray}
{\bf e_1} &=& \bar{a} (0,1), {\bf e_2} = \bar{a}/2(\sqrt{3},-1), 
{\bf e_3} = \bar{a}/2 (-\sqrt{3},-1),\nonumber \\ 
{\bf a_1} &=& \bar{a}/2 (\sqrt{3},3), {\bf a_2} = \bar{a}/2 (-\sqrt{3},3) 
\label{lattice-vector}
\end{eqnarray}
with $\bar{a}$ being the lattice constant between nearest-neighbor $A$ and $B$. 
 The spin-orbit coupling term is represented by 
the imaginary NNN hopping $\lambda_{SO}$ term within the same sublattice 
where $\nu_{ij}= 1$ for ${i,j}\in A$ (red counterclockwise 
arrows in Fig.~\ref{ribbon}) and 
$\nu_{ij}=-1$ for ${i,j}\in B$ (blue clockwise arrows in Fig.~\ref{ribbon}).
\begin{figure}[t]
\begin{center}
%\vspace{0.2cm}
\includegraphics[width=8.5cm,clip]{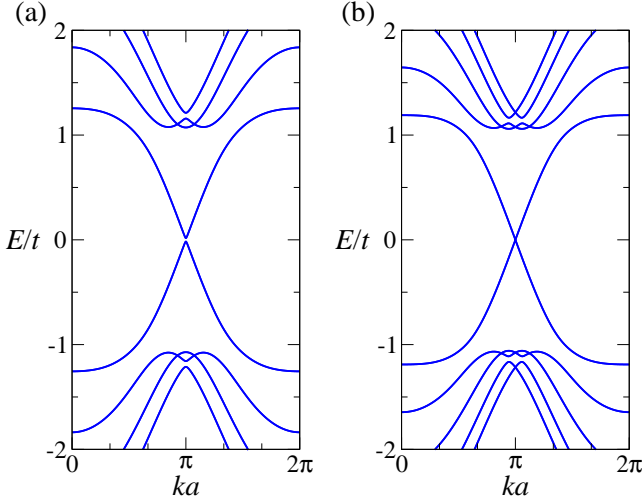}
\end{center}
\par
 \vskip -0.7cm
\caption{
(Color online)
Energy spectrum of a finite-sized Kane-Mele model on a zigzag ribbon 
for (a) $N=4$, (b) $N=5$  
of honeycomb lattice. Here, we set $t=1$, $\lambda_{SO}/t =0.2$.}
\label{spectrum}
\end{figure}
In the absence of the SO coupling, the KM model on zigzag ribbon 
reduces to the tight-binding 
Hamiltonian of a 2D zigzag graphene nano-ribbon (ZGNR)\cite{kyoko}, 
which shows two in-equivalent Dirac points located 
 at $k\equiv k_x = \pm \frac{2\pi}{3a}$ with $k_x$ being 
momentum along $x-$axis with $a\equiv \sqrt{3}\bar{a}$. 
Meanwhile, there exists a zero-energy flat band extended in the interval of 
$2\pi/3 \le ka \le 4\pi/3$, known to correspond to the edge state of ZGNR\cite{kyoko,fujita}.  It has been shown that 
the magnitudes of the 
edge state wave functions decay exponentially with distance away from the  
two edges, and the edge states are completely 
localized at the edges for $ka=\pi$\cite{fertig,louie}.

In the presence of SO couping, the KM Hamiltonian $H_{KM}$ 
for a finite-sized zigzag ribbon 
(see Fig.~\ref{ribbon}) on honeycomb 
lattice supports helical edge states $\Psi_{R,1(2)}^{\uparrow (\downarrow)}, 
\Psi_{L,1(2)}^{\downarrow (\uparrow)}$ with topological nature\cite{KM,hur}. 
Here,  
$\Psi_{R,1(2)}^{\uparrow(\downarrow)}$ stands for the wave function of 
the right-moving 
edge state electron with spin up (spin down) along the edge $1$ ($2$), 
respectively. The indices $1$ and $2$ also refer to the top 
and bottom edge, respectively. 
Similarly, $\Psi_{L,1(2)}^{\downarrow\uparrow}$ stands for the the wave function 
of the left-moving 
edge state electron with spin down (spin up) along the edge $1$ ($2$), 
respectively. The helical nature of these topological edge states 
manifest itself in the lock-in between the electron spin configuration 
and the direction of its momentum. 

In the limit of large ribbon size $N\gg 1$, 
the electron operator $c_{i}^\sigma (x)$ 
near the edge can be decomposed approximately in terms of these well-localized 
edge states as:
\begin{eqnarray}
c_{1(2)}^{\uparrow (\downarrow)}(x) &\approx& \Psi_{R,1(2)}^{\uparrow (\downarrow)}(x) 
e^{{\it i} k_F x},\nonumber \\ 
c_{1(2)}^{\downarrow (\uparrow)}(x)&\approx& \Psi_{L,1(2)}^{\downarrow (\uparrow)}(x) e^{-{\it i} k_F x}.
\end{eqnarray}
\begin{figure}[t]
\begin{center}
%\vspace{0.2cm}
\includegraphics[width=8.5cm,clip]{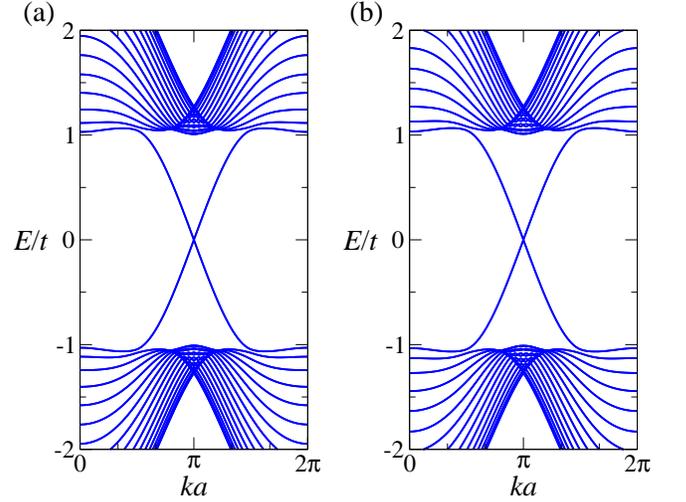}
\end{center}
\par
 \vskip -0.7cm
\caption{
(Color online)
Energy spectrum of a finite-sized Kane-Mele model on a zigzag ribbon 
for (a) $N=16$, (b) $N=15$  
of honeycomb lattice. Here, we set $t=1$, $\lambda_{SO}/t =0.2$.}
\label{spectrum2}
\end{figure}
The Hamiltonian of the edge $H_{edge}$ is therefore given by:
\begin{eqnarray}
H_{edge} &=& -{\it i} v_F \int dx  
[\Psi_{R,1}^{\dagger \uparrow} \partial_x \Psi_{R,1}^{\uparrow} - 
 \Psi_{L,1}^{\dagger \downarrow} \partial_x \Psi_{L,1}^{\downarrow}\nonumber \\
&+& 
\Psi_{R,2}^{\dagger \downarrow} \partial_x \Psi_{R,2}^{\downarrow} - 
 \Psi_{L,2}^{\dagger \uparrow} \partial_x \Psi_{L,2}^{\uparrow}
]
\label{H-edge}
\end{eqnarray}
with $v_F$ being the Fermi velocity.

At a finite system size, however, the edge state electron wave functions 
acquire an additional functional dependence on $y-$axis 
($c_{1(2)}^{\uparrow (\downarrow)}(x,y)$) and 
are found  to extend 
over a finite range in bulk via 
diagonalizing the tight-binding KM ribbon. The Hamiltonian of the 
edge states in this case are given by:
\begin{eqnarray}
H_{edge} &=& v_F \int dk \int dy k 
[\bar{\Psi}_{R,1}^{\dagger \uparrow}(k,y) 
\bar{\Psi}_{R,1}^{\uparrow}(k,y) \nonumber \\
&-& 
 \bar{\Psi}_{L,1}^{\dagger \downarrow}(k,y) 
 \bar{\Psi}_{L,2}^{\downarrow}(k,y)\nonumber \\
&+& 
\bar{\Psi}_{R,2}^{\dagger \downarrow}(k,y) 
\bar{\Psi}_{R,2}^{\downarrow}(k,y) - 
 \bar{\Psi}_{L,2}^{\dagger \uparrow}(k,y) \bar{\Psi}_{L,2}^{\uparrow}(k,y)
], \nonumber \\
\label{H-edge-finite}
\end{eqnarray}
where $\bar{\Psi}_{R/L,1(2)}^{\uparrow (\downarrow)}(k,y)$ are 
the edge state electron operators for a KM ribbon at a given 
momentum $k$ and $y$ obtained via Fourier transforming 
$c_{1(2)}^{\uparrow (\downarrow)}(x,y)$ along the $x-$axis:
\begin{eqnarray}
\bar{\Psi}_{R,1(2)}^{\uparrow (\downarrow)}(k,y) &=& 
\int dx e^{-{\it i}k x} 
c_{1(2)}^{\uparrow (\downarrow)}(x,y),\nonumber \\ 
\bar{\Psi}_{L,1(2)}^{\downarrow (\uparrow)}(k,y) &=& 
\int dx e^{-{\it i}k x} 
c_{1(2)}^{\downarrow (\uparrow)}(x,y).
\label{Psi-edge-finite}
\end{eqnarray}
Note that $\bar{\Psi}_{R/L,1(2)}^{\uparrow (\downarrow)}(k,y)$ 
can be obtained numerically as the eigenstates of the Dirac dispersed 
helical edge states via diagonalizing the finite-sized zigzag KM ribbon.  
As shown in Fig.~\ref{spectrum} and Fig.~\ref{spectrum2}, 
we numerically diagonalize the KM model at 
$N=even$ ($N=4,16$) and $N=odd$ ($N=5,15$) zigzag ribbon\cite{hur,van}. 
Two pairs of Dirac dispersed edge states  
($\bar{\Psi}_{R,1(2)}^{\uparrow (\downarrow)}, 
\bar{\Psi}_{L,1(2)}^{\downarrow (\uparrow)}$) emerge in 
the energy spectrum of a finite-sized 
KM zigzag ribbon, and they tend to intersect at the Dirac points $ka=\pm \pi$. 
However, at the Dirac points, a finite energy gap 
is developed for $N=even$, while no gap is seen for all $N=odd$ 
(see Fig.~\ref{spectrum}). We shall focus on this even-odd effect in more 
details below. 
%We identify the edge state wave functions 
%via eigenstates of the finite-sized zigzag KM ribbon. 
Similar to the case for ZGNR, for $2\pi/3 \le ka\le 4\pi/3$, 
we find the square magnitude of the 
two degenerate edge state eigenfunctions 
$|\Psi(y)|^2=|\bar{\Psi}_{L/R,i}(k,y)|^2$ 
(except for $N=even$ and $ka =\pm \pi$) show  
a symmetrical exponential decay from one edge to the other 
with respect to the ribbon center 
($y=L/2$) from both edges into the 
bulk as a function of the distance to the corresponding edge. 
Here, $y$ measures the distance to the edge along 
$y-$axis and $y=0$ corresponds to the first (top) zigzag chain. 
Also, to simplify the discussions, 
we use an integer index 
$y/b + 1 = N_i=1,2,\cdots N$ with $y=(N_i-1) b$ 
for labeling the $N_i$-th 
zigzag chain along $y-$axis for a ribbon with $N$ zigzag chains; 
$y=2b$ corresponds to the position of the 
third ($N_i=3$) zigzag chain. 
As shown in Fig.~\ref{decay-edge-Leven} (b)  
and Fig.~\ref{decay-edge-Lodd}, 
the decay of these edge states 
is well fitted 
by the following exponential form:
\begin{equation}
|\bar{\Psi}_{L/R,i}(k,y)|^2 \propto e^{-\beta y/b}
\label{wf-decay}
\end{equation} 
where $\beta$ is the decay constant depends on the momentum $k$. 
For $N=even$ and at the Dirac point $ka =\pi$, we find 
the right and left moving edge states get hybridized so that 
the square magnitudes $|\Psi(y)|^2 = |\bar{\Psi}_{hy,i}(y)|^2$ 
of the two degenerate edge states 
are maximized on both edges 
(see Fig.~\ref{decay-edge-Leven} (a)). 
Note that we find via eigenvector analysis of our numerical results   
through exact diagonalization of the finite-sized KM ribbon 
that these distinct two hybridized edge state wave-functions 
:$\bar{\Psi}_{hy,1}(y)\neq \bar{\Psi}_{hy,2}(y)$ show the same magnitudes: 
$|\bar{\Psi}_{hy,1}(y)|= |\bar{\Psi}_{hy,2}(y)|$.  
Numerically, the values of $|\Psi(y)|^2$ as a function of $y$ 
for a given edge state are obtained approximately 
by summing over the square of the matrix elements of the corresponding 
edge-state eigenvector contributed 
from both sublattices: 
$|\Psi(y)|^2 = |\Psi_A(y)|^2 + |\Psi_B(y+\bar{a}/2)|^2$. 
Also, we find the square magnitude $|\Psi(y)|^2$ at $ka=\pi$ 
for $N=even$ (see Fig.~\ref{decay-edge-Leven}(a)) 
oscillate along $y-$axis. Similar oscillations are found for $N=odd$ 
but not shown in Fig.~\ref{decay-edge-Lodd}(a) as 
the values of $|\Psi(y)|^2$ for $N=odd$ near edges 
are vanishingly small and go beyond the logarithmic scale shown there. 
This oscillatory 
behavior agrees qualitatively with that shown in Ref.~\onlinecite{sandler}.

\begin{figure}[t]
\begin{center}
%\vspace{0.2cm}
\includegraphics[width=5.0cm,clip]{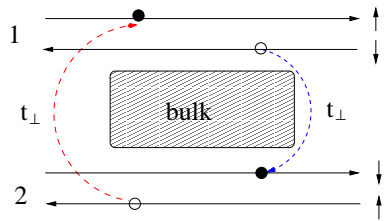}
\end{center}
\par
 \vskip -0.7cm
\caption{
(Color online)
Schematic diagram for the inter-edge hopping term $t_\perp$ 
(red or blue dashed line).}
\label{tperp-dia}
\end{figure}

Based on our numerical results, the edge states are much more 
localized at the Dirac point 
$ka=\pm \pi$: $\beta(k=\pi / a)> 1$ compared to that at other 
values of $k$. 
For $2\pi/3 <ka<\pi$, 
however, the edge state wave functions extend 
over a finite region in the bulk (see Fig.~\ref{decay-edge-Leven} (b)). 
In both cases, a weak but finite overlap between edge and bulk 
electron wave functions 
is expected to be present in the zigzag KM ribbon, 
which generates an effective inter-edge hoping $t_\perp$ term 
 approximately as (see Fig.~\ref{tperp-dia} and Sec. II B):
\begin{figure}[t]
\begin{center}
%\vspace{0.2cm}
\includegraphics[width=0.9\linewidth,clip]{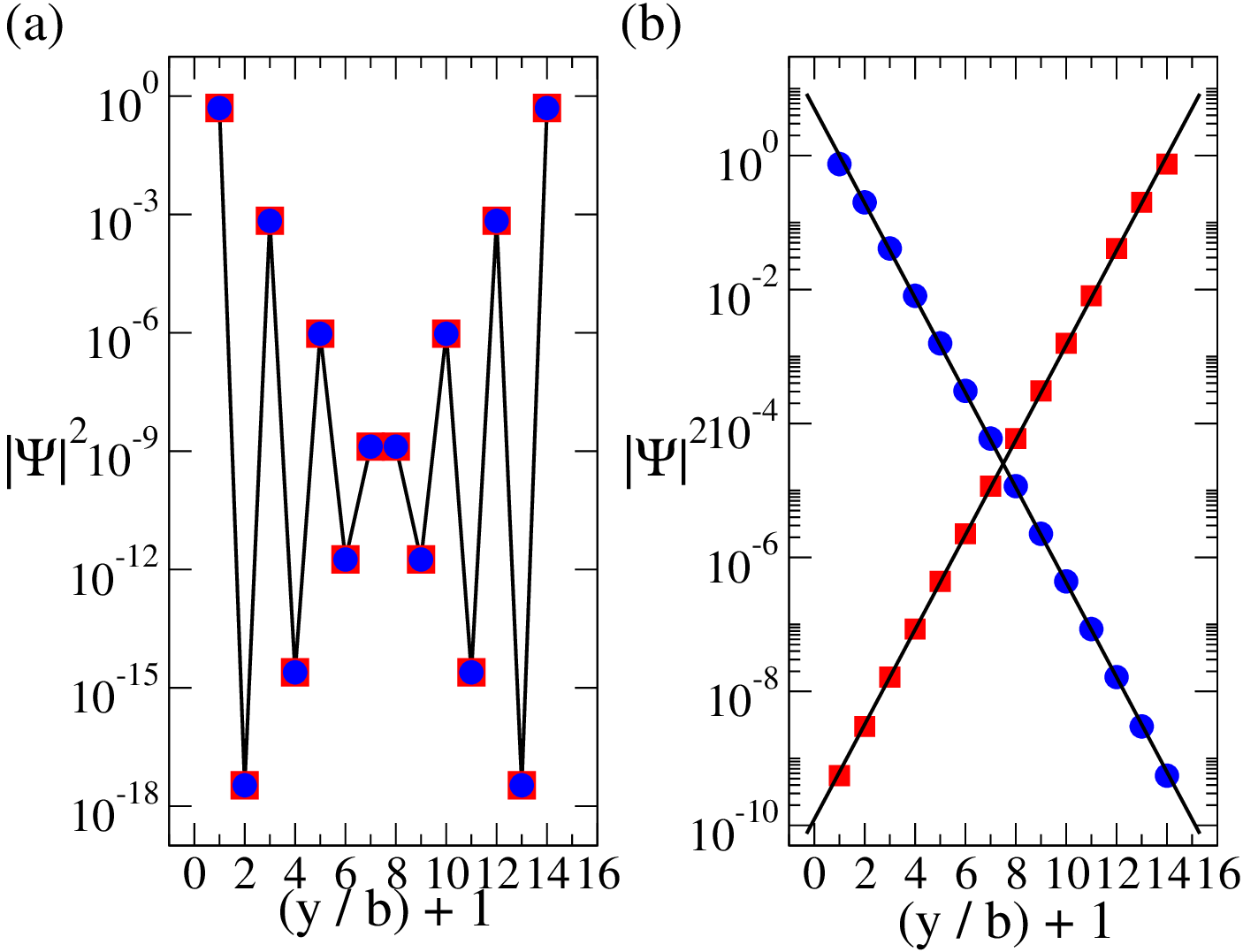}
\end{center}
\par
 \vskip -0.7cm
\caption{
(Color online)
The square magnitude of the 
edge state wave function $|\Psi|^2$ of the KM zigzag ribbon 
at half filling as a function of  
$y/b +1$ (defined in text) for $N=14$ and (a) for $ka=\pi$ and (b) 
for $ka=\pi\pm 0.2\pi$. Here, $|\Psi|^2$ 
(blue circles and red squares) represents for the square 
magnitude of the two edge state wave functions, which are 
degenerate eigenstates at the corresponding wave vector $k$. 
In (a), the two hybridized degenerate edge state wave functions 
$\Psi = \Psi_{hyb,i=1,2}$ 
(red and blue symbols) lead to the same square magnitude, 
$|\Psi_{hyb,1}|^2=|\Psi_{hyb,2}|^2$, in  
(b), we make the following identifications: 
$\Psi(y) = \Psi_{R,1}^{\uparrow}$ (blue) and 
$\Psi(y) = \Psi_{L,2}^{\uparrow}$ (red).  
The solid lines are guides to the eyes in (a), and in (b) 
they are fits to the exponential form in Eq.~(\ref{wf-decay}). 
We set $\lambda / t = 0.1$.
}
\label{decay-edge-Leven}
\end{figure}
\begin{eqnarray}
H_{t_\perp}&=& t_\perp \sum_{\sigma=\uparrow,\downarrow} 
\int dx [c^{\dagger\sigma}_1 c_2^{\sigma} + h.c.]\nonumber \\
&\approx& t_\perp  
\int dx e^{2{\it i}k_F x} 
\left(\Psi_{R,1}^{\dagger\uparrow} \Psi_{L,2}^\uparrow + 
\Psi_{R,2}^{\dagger\downarrow} \Psi_{L,1}^\downarrow\right)
+ h.c..\nonumber \\
\label{tperp}
\end{eqnarray}
with $x=na$ and $n=\pm 1,\pm 2,\cdots$. The value of 
$t_\perp$ in Eq.~(\ref{tperp}) can 
be estimated numerically via diagonalizing the finite-sized 
KM ribbon:
\begin{eqnarray}
H_{t_\perp}&=& t_\perp \sum_{\sigma=\uparrow,\downarrow} 
\int dx \int dy [c^{\dagger\sigma}_1(x,y) c_2^{\sigma}(x,y) + h.c.]\nonumber \\
&\approx& t_\perp  
\int dy 
[\bar{\Psi}_{R,1}^{\dagger\uparrow}(k_F,y) 
\bar{\Psi}_{L,2}^\uparrow(k_F,y)\nonumber \\
&+&\bar{\Psi}_{R,2}^{\dagger\downarrow}(k_F,y) 
\bar{\Psi}_{L,1}^\downarrow(k_F,y)]
+ h.c..\nonumber \\
\label{tperp-ribbon}
\end{eqnarray}
The $H_{t_\perp}$ turns out 
to be important in our RG analysis on the stability 
of the helical edge states (see below). 
The magnitude of $t_\perp$ can be estimated 
via the overlap integral\cite{mermin} 
of the opposite edge state wave functions through 
exact diagonalization of the tight-binding KM model at a 
finite-sized ribbon (see Eq.~(\ref{tperp-ribbon}))\cite{der-hau-lee}: 
\begin{equation}
t_\perp \approx t \int_0^L dy [\bar{\Psi}_{R,1}^{\ast\uparrow}(y) 
\bar{\Psi}_{L,2}^{\uparrow}(y)  +  \bar{\Psi}_{L,1}^{\ast\downarrow}(y) 
\bar{\Psi}_{R,2}^{\downarrow}(y) + c.c.],  
\label{tperp-overlap}
\end{equation}
where we have dropped the $k_F$ dependence in 
$\bar{\Psi}_{L/R,\alpha}^{\sigma}(k_F,y)$ in Eq.~(\ref{tperp-overlap}). 
At half filling, $k_F a =\pm \pi$, hence $e^{2{\it i} k_F x} = 1$ and 
$H_{t_\perp}$ can in general survive. However, $N=even$ and $N=odd$ 
lead to different results in this case as explained below.  

For $N=even$, due to breaking of the sublattice translational invariance 
at the boundaries results in a finite  
$t_\perp$. This leads to opening up a gap $\Delta$ in the excitation 
spectrum at the Dirac point 
when combining Eqs.~(\ref{H-edge}) and (\ref{tperp}):
\begin{eqnarray}
\epsilon (k -\pi /a) &\approx& \pm \sqrt{v_F^2 (k-\pi/a)^2 + (\Delta / 2) ^2}
\end{eqnarray} 
with $\Delta = 2t_\perp$. We numerically analyzed the gap $\Delta$ 
as shown in Fig.~\ref{gap}. 
The existence of a finite $t_\perp$ not only agrees with  
the energy gap at the Dirac point, it also explains the hybridization 
of the left and right moving edge states that we found in numerics as 
the eigenstates of the edge states in the presence of $t_\perp$ are 
linear combinations of left and right moving edge states. 
It is clear from Fig.~\ref{gap}(a) that the magnitude of the 
gap decreases with increasing the ribbon size $L$. In fact, it shows 
an exponential decay (see Fig.~\ref{gap} (b)): 
\begin{figure}[t]
\begin{center}
%\vspace{0.2cm}
\includegraphics[width=0.9\linewidth,clip]{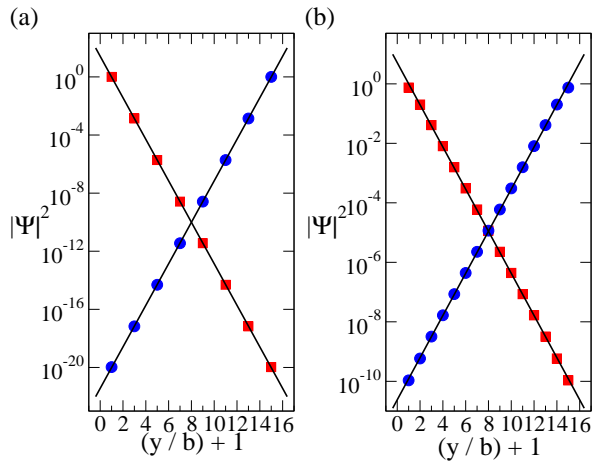}
\end{center}
\par
 \vskip -0.7cm
\caption{
(Color online)
The square magnitude of the 
edge state wave function $|\Psi|^2$ of the KM zigzag ribbon 
at half filling as a function of  
$y/b + 1$ for (a) $N=15$ and $ka=\pi$ and (b) for $N=15$ and 
$ka=\pi\pm 0.2\pi$. Here, $|\Psi|^2$ 
(blue circles and red squares) represents for the square 
magnitude of the two edge state wave functions, which are 
degenerate eigenstates at the corresponding wave vector $k$. 
The solid lines are fits to the exponential form in Eq.~(\ref{wf-decay}). 
We set $\lambda / t = 0.1$. Note that in (a) $|\Psi|^2$ is shown for  
only even values of $y/b$ (see text).
}
\label{decay-edge-Lodd}
\end{figure}
\begin{equation}
\Delta \approx \Delta_0 e^{-\alpha L}
\end{equation}  
with $\alpha$ being the decay constant.

Note that the decay of the small gap $\Delta$ 
was found to be power-law fashion 
in Ref.~\onlinecite{sandler} by a different (analytical) approach based on 
the analytical eigenstates for KM model on 
2D honeycomb lattice. 
With increasing $\lambda_{SO}$, 
we find the magnitude of $\Delta$ increases with increasing $\lambda_{SO}$, 
which comes as a result of the increase in bulk band gap $\Delta_{SO}$.  
We will show in Sec. IV. that 
this gaped phase corresponds to the charge and spin insulating (or II) phase. 
In the limit of infinite ribbon 
width $L\rightarrow \infty$, the gap $\Delta$ vanishes and the gapless 
Dirac spectrum is recovered. However, for $N=odd$, the sublattice 
translational symmetry at boundaries leads to cancellations in the overlap 
integral Eq.~(\ref{tperp-overlap}) between sublattices $A$ and $B$. 

At a generic filling away from half-filled, 
the oscillatory phase factor $e^{2{\it i}k_F x}$ in $t_\perp$  term 
results in cancellations upon averaging over $x$ and $H_{t_\perp}$ hence 
vanishes. As shown below, we also numerically 
confirmed this result via Eq.~(\ref{tperp-overlap}).
Though $H_{t_\perp}$ term survives only for $N=even$ and at half filling, 
as shown below, additional two-particle 
scattering terms are generated via second-order inter-edge tunnelings, 
which play an important role in all above-mentioned cases 
in our stability analysis 
of the helical edge states in KMH ribbon.

\subsection{The Kane-Mele Hubbard model on a zigzag ribbon}

Based on the above results for the non-interacting KM model on a 
finite-sized zigzag ribbon, we now perform an analytical analysis via 
perturbative RG approach on the weakly 
interacting KM model (the KM Hubbard model) by including
 a weak on-site Hubbard $U$ term in $H_{KM}$. 
Upon including the on-site Hubbard $U$ term, the Hamiltonian of the 
Kane-Mele-Hubbard (KMH) model reads:
\begin{eqnarray}
H_{KMH} &=& H_{KM} +H_U,\nonumber \\
H_U &=& U\int dx \int dy [n^{\uparrow}(x,y) n^{\downarrow}(x,y)],\nonumber \\
n^\sigma(x,y)  &=& c^{\dagger\sigma}(x,y) c^{\sigma}(x,y)
\label{H_KMH}
\end{eqnarray}
with $U>0$. 
To simplify our calculations, we consider $H_{KM}$ approximately as 
three different contributions: (i) 
the well-localized edge state $H_{edge}$, (ii) 
the insulating bulk states $H_b$, and (iii) a weak coupling between 
edge and the bulk states $H_{t^\prime}$ due to the finite-size effect: 
\begin{equation}
H_{KM} \approx H_{edge} + H_b + H_{t^\prime},
\label{HKM-size}
\end{equation}
where the edge part $H_{edge}$ is defined in Eq.~(\ref{H-edge}), 
the bulk part $H_b$ of $H_{KM}$ is given by:
\begin{eqnarray} 
H_{b} &=& \sum_{k,\alpha=\uparrow,\downarrow} 
H_{KM}(c_b^\alpha (k), c_b^{\dagger,\alpha}(k)), \nonumber\\ 
\label{Hb}
\end{eqnarray}
and the edge-bulk overlap term $H_{t^\prime}$ reads:
\begin{eqnarray} 
H_{t^{\prime}} &=&  t^\prime
\int dx  [ e^{-{\it i} k_F x } \Psi_{R,1}^{\dagger \uparrow} c_{b,1}^\uparrow (x) 
 + e^{{\it i} k_F x } \Psi_{L,1}^{\dagger \downarrow} c_{b,1}^\downarrow (x) \nonumber \\
&+& 
 e^{-{\it i} k_F x } \Psi_{R,2}^{\dagger \downarrow} c_{b,2}^\downarrow (x) 
 + e^{{\it i} k_F x } \Psi_{L,2}^{\dagger \uparrow} c_{b,2}^\uparrow (x)], \nonumber \\
\label{Htprime}
\end{eqnarray}
where $t^\prime \sim \mathcal{O}(t,\lambda_{SO})$. 
\begin{figure}[t]
\begin{center}
%\vspace{0.2cm}
\includegraphics[width=8.5cm,clip]{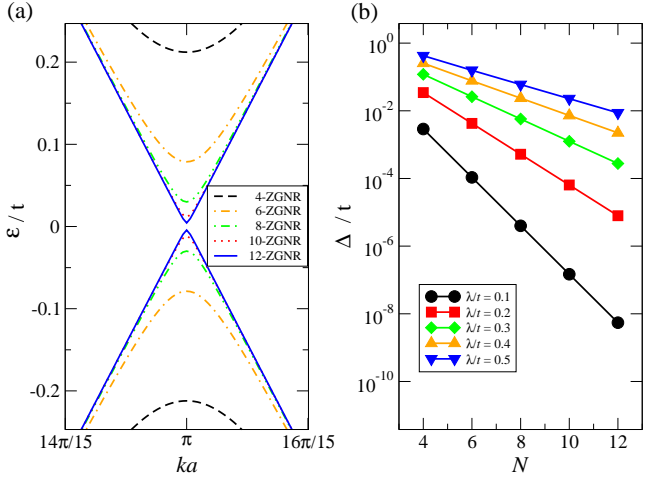}
\end{center}
\par
 \vskip -0.7cm
\caption{
(Color online)
(a) Energy spectrum $\epsilon$  
versus momentum $k$ 
of the topological edge states 
of a finite-sized ($N$ zigzag chains) 
Kane-Mele model on a zigzag ribbon 
of honeycomb lattice near the Dirac point $ka=\pi$ for different 
ribbon sizes. Here, we set $t=1$, $\lambda_{SO}/t =0.5$. 
(b) Energy gap $\Delta$ at the Dirac point 
as a function of $N$ for 
different values of $\lambda$.}
\label{gap}
\end{figure}
where $c_{b,1(2)}^\sigma (x)$ stands for the bulk electron operators near 
edge $1(2)$. 
We further simplify the Hubbard $U$ term $H_U$ in Eq.~(\ref{H_KMH}), and 
decompose it into the edge $H_{U,e}$ 
and the bulk $H_{U,b}$ contributions as:
\begin{eqnarray}
H_U &=& H_{U,e} + H_{U,b}, \nonumber \\
H_{U,e} &=& U \int dx 
 \sum_{i=1,2}[n_{i}^{\uparrow}(x) n_{i}^{\downarrow}(x)] , \nonumber \\
H_{U,b} &=& U \int dx \int dy [n_{b}^{\uparrow}(x,y) n_{b}^{\downarrow}(x,y)].
\end{eqnarray}
Here, $i=1 (2)$ refers to the top (bottom) edge, $c_b^\alpha(k)$ is the 
electron destruction operator in the bulk. 
Also, the $H_b$ term, representing the KM 
model of the bulk electrons, shows an energy dispersion $E_b(k)$ 
with an energy gap 
$\Delta_{so} \sim 6\sqrt{3} \lambda_{SO}$\cite{hur}. 
For the periodic 2D KM model, 
$E_b(k)$ has been shown to be (see Ref.~\onlinecite{hur}):
\begin{eqnarray}
E_b(k) &=& \pm \sqrt{|g_k|^2 +\gamma_k^2}, \nonumber\\
g_k &=& t \sqrt{3+2\cos(\sqrt{3}k_y) + 4 \cos(\sqrt{3}k_y/2) \cos(3k_x/2)}, \nonumber \\
\gamma_k &=& \lambda_{SO}
[-\sin(\sqrt{3}k_y) + 2\cos(3k_x/2) \sin(\sqrt{3}k_y/2)]. \nonumber \\
\end{eqnarray}
%We will focus here on the stability of the topological edge 
%states upon including 
%a weak on-site Coulomb interaction $U$ (the Hubbard term $H_U$) as well as 
%the inter-edge hopping $t_\perp$ term. 
To simplify our analysis, we assume here the bulk bands 
are well-separated by the bulk gap $\Delta_{SO}$ in the presence of a finite spin-orbit coupling $\lambda_{SO}$, and $U\ll \lambda_{SO}$. 
The on-site Hubbard $U$ term along the edges $H_{U,e}$ can be re-written  as:
\begin{eqnarray}
H_{U,e} &=& H_{\rho} + H_\sigma^z,\nonumber \\
H_\rho &=& g_\rho \int dx J_{L}^\rho J_{R}^\rho, \nonumber \\
H_\sigma^z &=&  g_\sigma^z \int dx \vec{J}_L^z \vec{J}_R^z, \nonumber \\
%H_{\delta v_F} &=& \frac{\pi}{2} \delta v_F^c  \int dx \sum_{i=1,2} (J^\rho_{L,i}J^\rho_{L,i} + J^\rho_{R,i}J^\rho_{R,i}),
%\nonumber \\
\label{H-U-edge}
\end{eqnarray}
where $J^\rho_{R/L}$ is the $U(1)$ scalar current operator 
and $J^{z}_{L/R}$ is the $z-$component 
of the $SU(2)$ vector current operator $\vec{J}^{a=x,y,z}_{L(R)}$,  
defined respectively 
as\cite{gogolin,giamarchi}: 
\begin{eqnarray}
J^\rho_{L(R)} &=& \sum_{i=1,2} J^\rho_{L(R),i},\nonumber \\
J^\rho_{L,1(2)} &=& \Psi_{L,1(2)}^{\dagger\downarrow(\uparrow)} 
\Psi_{L,1(2)}^{\downarrow(\uparrow)}, \nonumber \\
J^\rho_{R,1(2)} &=& \Psi_{R,1(2)}^{\dagger\uparrow(\downarrow)} 
\Psi_{R,1(2)}^{\uparrow(\downarrow)}, \nonumber \\
\vec{J}^{a=x,y,z}_{L(R)} &=& 
\Psi^{\dagger\alpha}_{L(R)}\vec{\sigma}^a_{\alpha\beta}\Psi_{L(R)}^\beta,\nonumber\\
\vec{J}_{L(R)}^z &=& \frac{1}{2} (\Psi_{L,2 (R,1)}^{\dagger\uparrow} 
\Psi_{L,2 (R,1)}^{\uparrow} - \Psi_{L,1 (R,2)}^{\dagger\downarrow} 
\Psi_{L,1 (R,2)}^{\downarrow}).\nonumber\\
\label{J-current}
\end{eqnarray}
Here, $g_\rho$ and $g_\sigma^z$ take the following 
bare (initial) values in the context of renormalization group 
analysis: $g_\rho (\mu_0) \equiv g_{\rho,0} = U/2$, $g_\sigma^z (\mu_0) \equiv g_\sigma^{z,0}= -2U$ with $\mu_0$ being the bandwidth of the tight-binding 
KM model. 
%Also, $\delta v_F^c$ term will only renormalizes the 
%Fermi velocity $v_F^c$ of the edge state electrons in the charge sector 
%(see below): 
%$v_F^c \rightarrow v_F^c + \frac{2}{\pi}\delta v_F^c$ with $\delta v_F^c = -U$.
\begin{figure}[t]
\begin{center}
%\vspace{0.2cm}
\includegraphics[width=8.5cm,clip]{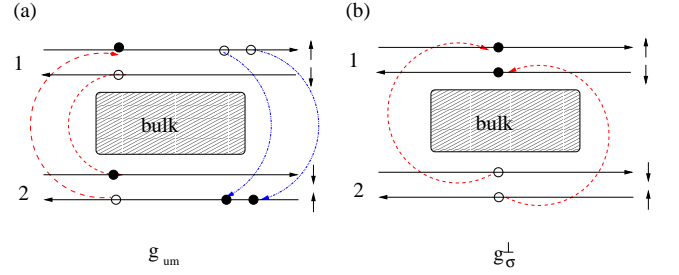}
\end{center}
\par
 \vskip -0.7cm
\caption{
(Color online)
Schematic diagrams for (a) the inter-edge Umklapp $g_{um}$ 
(red and blue arrows) and (b) the inter-edge 
spin-flip $g_\sigma^\perp$ processes.}
\label{gperpgz-dia}
\end{figure}

We now turn our attention to $H_{t^\prime}$ term in Eq.~(\ref{HKM-size}). 
Integrating out the bulk electron $c_b^\alpha$ 
in Eqs.~(\ref{Hb}) and ~(\ref{Htprime}), 
an effective inter-edge tunneling term $H_{t_\perp}$ as shown in 
Eq.~(\ref{tperp}) is generated where 
$t_\perp \sim D_{bulk} (t^{\prime})^2/ \Delta_{SO}$ with $D_{bulk}$ being 
the average electron density of states in the bulk. 
The estimation for $t_\perp$ here  
can be compared to that in Eq.~(\ref{tperp-overlap}) via numerical 
diagonalization of the KM ribbon. 
Note that the inter-edge hoping $t_\perp$ 
(or the bulk gap $\Delta_{SO}$) 
is enhanced with increasing spin-orbit 
coupling $\lambda_{SO}$: $t_\perp \propto (t^{\prime})^2/ \Delta_{SO} 
\propto \lambda_{SO}^2/\Delta_{SO}\propto \lambda_{SO}$ (see Fig.~\ref{gap}(b)). 
Apart from $H_{t_\perp}$, the linear term in $t_\perp$,  
for both half filling and away from half filling cases, 
$H_{t_\perp}$ term will generate through the second order perturbation theory 
the following two two-particle scattering terms which turn 
out to be important in the stability analysis of topological edge states:
\begin{eqnarray}
\tilde{H}_{t_\perp} &=& H_{um} + H_\sigma^\perp,\nonumber \\
H_{um} &=& g_{um} \int dx  [ e^{{\it i} 4 k_F x} 
[ \Psi_{R,1}^{\dagger\uparrow} \Psi_{R,2}^{\dagger\downarrow} 
\Psi^{\uparrow}_{L,2}\Psi^{\downarrow}_{L,1}\nonumber \\
&+& \frac{1}{2} (\Psi_{R,1}^{\dagger\uparrow}(x) \Psi_{R,1}^{\dagger\uparrow}(x+a) 
\Psi^{\uparrow}_{L,2}(x)\Psi^{\uparrow}_{L,2}(x+a) \nonumber \\
&+& \Psi_{R,2}^{\dagger\downarrow}(x) \Psi_{R,2}^{\dagger\downarrow}(x+a) 
\Psi^{\downarrow}_{L,1}(x)\Psi^{\downarrow}_{L,1}(x+a)
)] + h.c.],\nonumber \\
H_{\sigma}^\perp &=& g_\sigma^\perp \int dx (J_L^+ J_R^- + h.c.), 
\label{Htperp-2nd}
\end{eqnarray}
 where $H_{um}$ and $H_\sigma^\perp$ represent for 
the inter-edge Umklapp and inter-edge spin-flip terms, respectively 
(see Fig.~\ref{gperpgz-dia}), 
and the transverse components of the $SU(2)$ vector current operators 
$J^{+}_{L/R}$, $J^{-}_{L/R}$ are defined as:
\begin{eqnarray}
J^{+}_{L(R)} &\equiv& \vec{J}_{L(R)}^x +{\it i} \vec{J}_{L(R)}^y = \Psi_{L,2 (R,1)}^{\dagger\uparrow} 
\Psi_{L,1 (R,2)}^{\downarrow},\nonumber \\ 
J^{-}_{L(R)} &\equiv& \vec{J}_{L(R)}^x -{\it i} \vec{J}_{L(R)}^y = 
\Psi_{L,1 (R,2)}^{\dagger\downarrow} \Psi_{L,2 (R,1)}^{\uparrow},\nonumber \\ 
\label{J+-}
\end{eqnarray}
Similar to Eq.~(\ref{tperp-ribbon}), 
the bare couplings for $H_\sigma^\perp$ and $H_{um}$, 
$g_{um}(\mu_0) \equiv g_{um}^0$ and 
$g_\sigma^\perp(\mu_0) \equiv g_\sigma^{\perp,0}$ can be 
estimated numerically as:
\begin{eqnarray}
g_{um}^0 &\approx& \frac{t}{4} \int_0^L dy [ \Psi_{R,1}^{\ast\uparrow}(y) 
\Psi_{R,2}^{\ast\downarrow}(y) 
\Psi^{\uparrow}_{L,2}(y) \Psi^{\downarrow}_{L,1}(y)\nonumber \\
&+& \frac{1}{2} (\Psi_{R,1}^{\ast\uparrow}(y) \Psi_{R,1}^{\ast\uparrow}(y) 
\Psi^{\uparrow}_{L,2}(y)\Psi^{\uparrow}_{L,2}(y) \nonumber \\
&+& \Psi_{R,2}^{\ast\downarrow}(y) \Psi_{R,2}^{\ast\downarrow}(y) 
\Psi^{\downarrow}_{L,1}(y)\Psi^{\downarrow}_{L,1}(y)
) + c.c],\nonumber \\
g_\sigma^{\perp,0} &\approx& \frac{t}{2} \int_0^y dy 
 [\Psi_{L,2}^{\ast\uparrow}(y) 
\Psi_{L,1}^{\downarrow}(y) 
\Psi^{\ast\downarrow}_{R,2}(y) \Psi^{\uparrow}_{R,1}(y)+ c.c.].\nonumber \\
\label{bare-gum-gsigma}
\end{eqnarray}
\begin{figure}[t]
\begin{center}
%\vspace{0.2cm}
\includegraphics[width=7.0cm,clip]{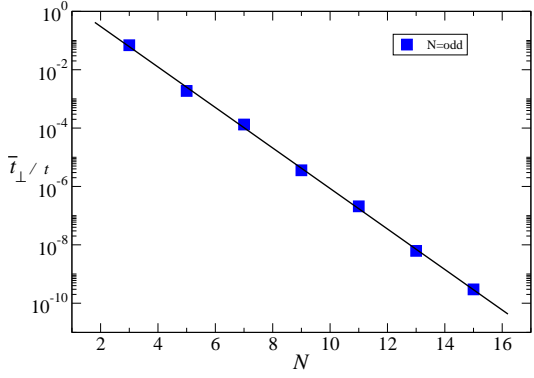}
\end{center}
\par
 \vskip -0.7cm
\caption{
(Color online)
The exponential decay of $\bar{t}_\perp$ as a function of odd 
number of zigzag chains $N$.}
\label{bartper}
\end{figure}

Note that the inter-edge Umklapp term $H_{um}$ 
depends sensitively 
on the electron filling factor.  
At half filling, $e^{{\it i}4k_F x} = 1$, $H_{um}$ therefore 
in general survives. For $N=even$, we find 
$-g_{um}^0= g_\sigma^{\perp,0}= t_\perp^2/t$ 
via the energy 
gap $\Delta$ at the Dirac point. 
For $N=odd$, by substituting the edge state wave functions 
that we numerically obtained based on the tight-binding KM ribbon 
into Eq.~(\ref{bare-gum-gsigma}), 
we find $-g_{um}^0= g_\sigma^{\perp,0}\equiv \bar{t}_\perp^2/t$, 
where
\begin{equation}
\bar{t}_\perp^2 \approx t^2 \int_0^L dy |\Psi_{R,1}^{\uparrow}(y)|^2 
|\Psi_{L,2}^{\uparrow}(y)|^2.
\end{equation} 
We further find numerically that $\bar{t}_\perp$ shows 
an exponential decay with increasing the ribbon width 
$L$, similar to the case for $N=even$:
\begin{equation}
\bar{t}_\perp \propto e^{-\gamma_k L}
\end{equation}
with $\gamma_k$ being the decay constant (see Fig.~\ref{bartper}). 
Note that at half filling, $\gamma_{k=\pi/a}\gg 1$ 
(or $t_\perp /t \ll 1$) due to the well-localized edge states. 

When the system is away 
from half filling, however, 
the oscillatory factor $e^{{\it i} 4 k_F x}$ 
in $H_{um}$ leads to cancellations upon summing over $x$, 
and therefore $H_{um}$ term vanishes completely. Nevertheless, 
$H_\sigma^\perp$ term still survive: 
$g_\sigma^{\perp,0}\equiv \bar{t}_\perp^2/t$. 

Note that similar two-particle scattering processes $H_\sigma^\perp$ and 
$H_{um}$ terms have been considered in Ref.~\onlinecite{Teo} in the 
context of the 
tunneling between helical edge states in a quantum point contact (QPC) 
as well as in Ref.~\onlinecite{fujimoto}. 
However, the authors in Ref.~\onlinecite{Teo} studied the effect 
of inter-edge single- and two-particle scattering processes on the 
helical edge states for a fixed electron-electron interactions 
(or Luttinger parameter $K$), while in Ref.~\onlinecite{fujimoto} 
the authors did not 
specify the origins of these two-particle scattering terms.  
By contrast, the two-particle scatterings we consider here come as a result of 
second-order inter-edge tunnelings. Furthermore, we 
treat the combined effects of the inter-edge 
two-particle scatterings $H_\sigma^\perp$, $H_{um}$ contributed from 
the inter-edge hopping $H_{t^\perp}$ as well as 
$H_\rho$, $H_\sigma^z$ terms 
via on-site Hubbard $U$ term in the weak-coupling limit on equal-footing.

Combining Eqs.~(\ref{H-U-edge})-(\ref{J+-}), 
the effective Hamiltonian of two  
weakly coupled helical edge states is therefore given by:
\begin{eqnarray}
H^{eff}_{edge} &=& H_{edge} + H_{U,e} + \tilde{H}_{t_\perp}\nonumber \\ 
             &=& H_{edge}+  H_{\sigma}^\perp + H_\sigma^z + H_\rho + H_{um}.\nonumber \\
\label{H_Hubbard}
\end{eqnarray}
where $H_{edge}$ can be re-expressed in terms of the scalar and vector current operators as, similar to that for an one-dimensional non-interacting 
electrons at half filling\cite{gogolin,giamarchi}:
\begin{eqnarray}
H_{edge} &=& \int dx [\frac{\pi}{2} v_F^c \sum_{i=1,2} (J^\rho_{L,i}J_{L,i}^\rho + J^\rho_{R,i}J_{R,i}^\rho)\nonumber \\
&+& \frac{2\pi}{3} v_F^s (\vec{J}_L \cdot \vec{J}_L + \vec{J}_R \cdot \vec{J}_R)]  
\end{eqnarray}
with the bare values for the Fermi velocities in the charge and spin sectors 
given by: $v_F^c = v_F^s = v_F$. 
%Meanwhile, $H_{edge}$ term gets corrections from $H_{U,e}$ term and 
%becomes $\tilde{H}_{edge}$, which has the same form as $H_{edge}$ with the 
%a correction in Fermi velocity in the charge sector: 
%$v_F^c \rightarrow \tilde{v}_F^c = v_F+\frac{2}{\pi}\delta v_F$, but no corrections in the spin sector: $v_F^s\rightarrow v_F^s$ (see Eq.~(\ref{H-U-edge})). 
Note that our effective 
Hamiltonian for the edges Eq.~(\ref{H_Hubbard}) describes two weakly coupled 
helical Luttinger liquids. In particular, $H_{edge}$, 
describing two non-interacting helical edge states, exhibits 
$U(1)\times SU(2)$ symmetry; while as 
$H_\sigma^\perp + H_\sigma^z$ in the couplings between two edges break 
the $SU(2)$ spin rotational symmetry down to $Z_2$ symmetry. 
Our effective model for the weakly-coupled helical Luttinger liquids 
$H_{edge}$ can be characterized as one-dimensional fermionic 
Hubbard model with $SU(2)$ spin-anisotropic interactions\cite{hur,gogolin,giamarchi}. 
The breaking of the $SU(2)$ symmetry of the model 
comes as a result of the Hubbard 
$U$ term at the edges (see Eq.~(\ref{H_KMH})).
\begin{figure}[t]
\begin{center}
%\vspace{0.2cm}
\includegraphics[width=7.5 cm,clip]{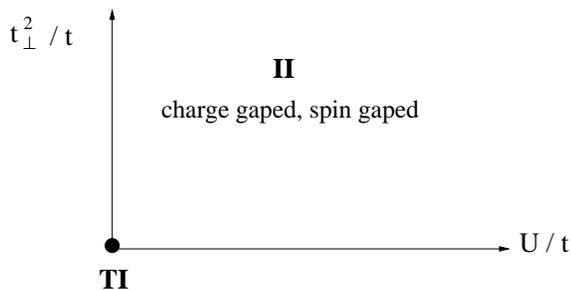}
\end{center}
\par
 \vskip -0.7cm
\caption{
(Color online)
Quantum phase diagram of the Kane-Mele Hubbard 
model at half filling for $N=even$ as a function of $U/t$ and $t_\perp^2/t$. 
The helical topological edge states (TI phase) is stable 
only at $U=t_\perp=0$ (dark circle). 
For a finite ribbon size, $t_\perp\neq 0$, the system flows to a charge 
and spin gaped (charge and spin insulating or II) phase.}
\label{Phase-II}
\end{figure}
\section{ RG analysis and phase diagram of the KMH model.}

We now analyze Eq. ~(\ref{H_Hubbard}) via renormalization group approach 
the stability of the edge states in the presence of Hubbard interactions. 
Note that the Hamiltonian Eq.~(\ref{H_Hubbard}) is closely related to the 
spin anisotropic Hubbard model for one-dimensional electrons where 
electron-electron interactions break the SU(2) 
symmetry\cite{gogolin,giamarchi}. 
Following the similar RG analysis to Refs.~\onlinecite{gogolin,giamarchi}, 
we may 
separate the four couplings $(g_\rho,g_{um},g_\sigma^\perp,g_\sigma^z)$ 
into two pairs belonging to the spin sector $(g_\sigma^z,g_\sigma^\perp)$ 
and the charge sector $(g_{um},g_\rho)$, respectively. Under RG 
transformations, these couplings exhibit the property of 
spin-charge separation, 
{\it i.e.} the 
renormalization of the couplings in the spin and charge sectors 
will remain in their own sector. We shall also analyze 
the single-particle inter-edge hopping 
$H_{t_\perp}$ term under RG.  
Below we separately discuss below 
the RG scaling equations for the half-filled and for a generic filling 
away from half filling for both $N=even$ and $N=odd$.
\begin{figure}[t]
\begin{center}
%\vspace{0.2cm}
\includegraphics[width=8.0cm,clip]{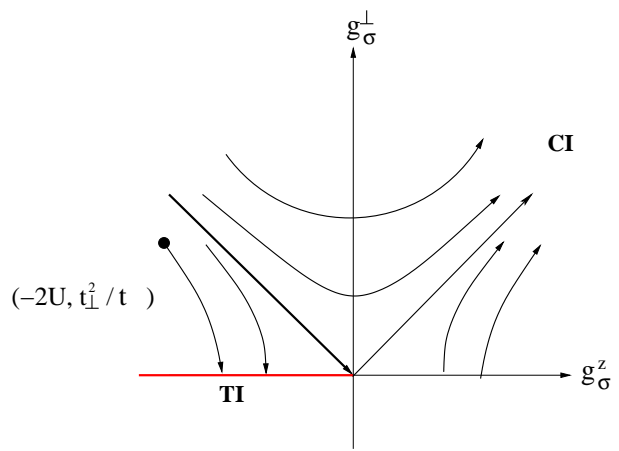}
\end{center}
\par
 \vskip -0.7cm
\caption{
(Color online)
The RG flows of the Kosterlitz-Touless type 
for the spin sector ($g_\sigma^\perp, g_\sigma^z$) 
of the zigzag Kane-Mele Hubbard ribbon for $N=even$ away from half filling. 
The black circle stands for the 
initial (bare) couplings. The arrows indicate the directions 
of the RG flows upon decreasing the curt-off scale $\mu$ from $\mu_0$. 
The red line represents a line of fixed points in the TI phase, 
the TI-CI phase boundary is defined by the separatrix line (thick black 
arrow). Note that the coupling $g_\rho$ does not flow under RG 
in this case (see Eq.~(\ref{RG-c-doped})).}
\label{RG-away-gpergz-even}
\end{figure}

\subsection{N=even}
\subsubsection{At half filling}

As shown previously, at half filling ($k_F a =\pm \pi$), the KM model for 
a finite-sized zigzag ribbon induces a finite inter-edge hopping term, 
$t_\perp \neq 0$. It can be shown that under RG 
transformation\cite{gogolin}, $H_{t_\perp}$ in Eq.~(\ref{tperp}) 
is a relevant operator with scaling dimension $[t_\perp]=-1$. 
Hence, the RG scaling equation reads\cite{gogolin}:
\begin{equation}
\frac{d t_\perp}{d \ln \mu} = -t_\perp,
\end{equation}
 where $\mu$ is the running cutoff in energy. 
Under RG transformation, the running cutoff scale $\mu$ is lowered 
from $\mu_0>0$ to zero. It is clear that $t_\perp$ flows to a strong 
coupling fixed point, $t_\perp(\mu=0)=\infty$. As a result, 
both $g_\sigma^\perp$ and $g_{um}$ become relevant under RG 
as their magnitudes are proportional to $t_\perp^2$. 
When the two-particle spin-flip processes $g_\sigma^\perp$ term becomes 
relevant, a spin gap is opening up, while a charge gap develops 
when the two-particle backscattering $g_{um}$ term becomes relevant. 
Therefore, the $t_\perp\rightarrow \infty$ fixed point 
corresponds to the charge and spin 
gaped (or charge and spin insulating II) phase (see Fig.~\ref{Phase-II}).

\begin{figure}[t]
\begin{center}
%\vspace{0.2cm}
\includegraphics[width=6 cm,clip]{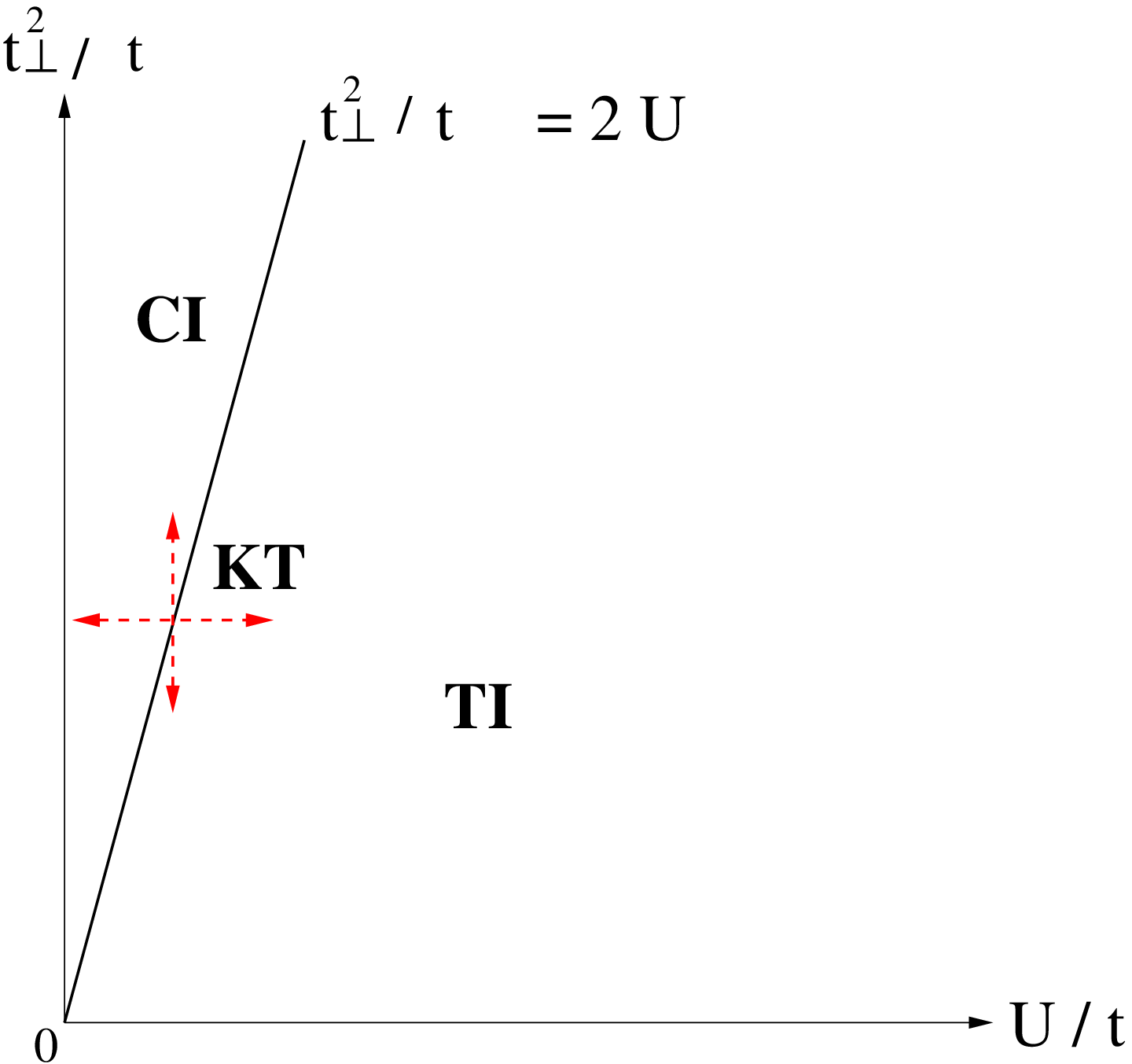}
\end{center}
\par
 \vskip -0.7cm
\caption{
(Color online)
Quantum phase diagram of the Kane-Mele Hubbard 
model away from half filling for $N=even$ 
as functions of $t_\perp^2/t$ and $U/t$. 
The helical topological edge states (TI phase) are unstable towards 
the charge conducting and spin insulating  
CI phase for $t_\perp^2/t>2U$. 
The TI-CI quantum phase transition set by the boundary 
$t_\perp^2/t=2U$ 
is of the Kosterlitz-Thouless (KT) type (red dashed arrows).}
\label{Phase-CI-even}
\end{figure}

\subsubsection{Away from half filling}

We now proceed to address the case of finite doping away from 
half filling, $k_F a \neq \pi$. 
In this case, the inter-edge hopping term $H_{t_\perp}$ and 
Umklapp term $H_{um}$ vanish due to the oscillatory exponential 
factors $e^{2{\it i}k_F x}$ and $e^{4{\it i}k_F x}$ respectively 
(see Sec. II). 
The RG scaling equations for both finite-sized and infinite-sized 
ribbons are reduced to\cite{gogolin,giamarchi}:
\begin{eqnarray}
\frac{d g_\rho}{d \ln \mu} &=& 0,\nonumber\\
\label{RG-c-doped-even}
\end{eqnarray}
in the charge sector with $g_\rho^0 = U$ and 
\begin{eqnarray}
\frac{d g_\sigma^\perp}{d \ln \mu} &=& -g_\sigma^\perp g_{\sigma}^z,\nonumber\\
\frac{d g_{\sigma}^z}{d \ln \mu} &=& -(g_\sigma^\perp)^2,\nonumber\\
\label{RG-s-even}
\end{eqnarray} 
in the spin sector with $(g_\sigma^{z,0}, g_\sigma^{\perp,0}) = 
(-2U, t_\perp^2/t)$. 

Via Eq.~(\ref{RG-c-doped-even}), it is clear that  
the system will not develop a charge gap under RG as 
$g_\rho$ does not diverge: $g_\rho(\mu)=g_\rho^0\ll 1$. 
The RG flows in the spin sector, however, suggest that the topological 
edge states may undergo the Kosterlitz-Thouless 
 transition upon increasing $t_\perp$ to a charge gapless but 
spin gaped (CI) phase characterized by the following fixed point:
\begin{eqnarray}
CI &:& g_\sigma^{\perp,0} + g_\sigma^{z,0} >0 , g_\sigma^{z}(\mu\rightarrow 0), g_\sigma^{\perp}(\mu\rightarrow 0) \rightarrow \infty,\nonumber \\
& & g_\rho(\mu\rightarrow 0)=0, g_{um}(\mu\rightarrow 0)=g_{um}^0\ll 1.\nonumber \\
& & \nonumber \\
\label{CI}
\end{eqnarray} 
 The TI-CI phase boundary is set by the separatrix  
$g_\sigma^\perp + g_\sigma^z=0$ (or when $t_\perp^2/t = 2U$, 
see Fig.~\ref{RG-away-gpergz-even}). The helical edge states are 
therefore stable 
for $t_\perp^2/t < 2U$, while it is unstable against the CI phase for $U< 
\frac{t_\perp^2}{2t}$. Combing RG flows in both charge and spin sectors, this spin gaped phase 
corresponds to the charge conducting but spin insulating (or CI) phase 
(see Fig.~\ref{Phase-CI-even}). 

\subsection{N=odd}

\subsubsection{At half filling}
\begin{figure}[t]
\begin{center}
%\vspace{0.2cm}
\includegraphics[width=7.5cm,clip]{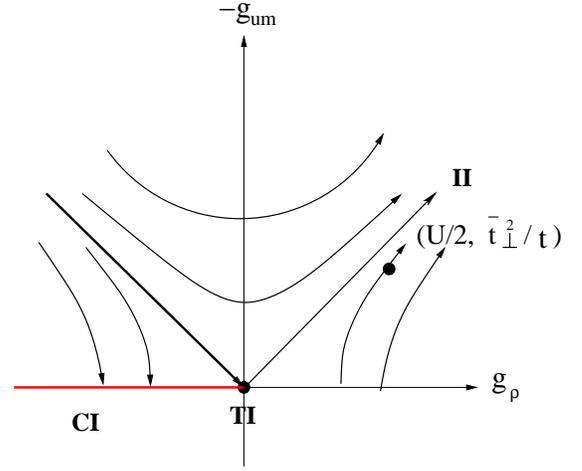}
\end{center}
\par
 \vskip -0.7cm
\caption{
(Color online)
The RG flows of the Kosterlitz-Touless type 
for the charge sector ($g_\rho, g_{um}$) 
of the zigzag Kane-Mele Hubbard ribbon at half filling for $N=odd$. 
The black circle stands for the 
initial (bare) couplings at 
$(g_{\rho}^0,-g_{um}^0) = (U/2,\bar{t}_\perp^2/t)$. 
The arrows indicate the directions 
of the RG flows upon decreasing the curt-off scale $\mu$ from $\mu_0$. 
The red line represents a line of fixed points in the CI phase, 
the CI-II  
boundary is defined by the separatrix line (thick black arrow) 
and its quantum transition is of the Kosterlitz-Thouless (KT) type. 
The topological TI phase is stable only at the origin $U=0=\bar{t}_\perp$.}
\label{RG-half-grhogum-odd}
\end{figure}
At half filling, $k_F a =\pm \pi$ and $t_\perp = 0$, all the four couplings 
$(g_\rho,g_{um},g_\sigma^\perp,g_\sigma^z)$ exist in general under RG 
transformations. Their  
initial (bare) couplings at $\mu =\mu_0$ are given by:  
$(g_{um}^0, g_{\rho}^0) = (-\bar{t}_\perp^2/t,U/2)$, 
$(g_\sigma^{z,0}, g_\sigma^{\perp,0}) = (-2U, \bar{t}_\perp^2/t)$.   
The RG scaling equations in this case 
can be casted in a spin-charge separated 
form\cite{gogolin,giamarchi} and are readily  
obtained via the operator product expansion (OPE) for the current algebra  
in the one-dimensional Hubbard model with broken $SU(2)$ symmetry 
(see, for example the Appendix in Chapter 17 of Ref.~\onlinecite{gogolin}):
\begin{eqnarray}
\frac{d g_\rho}{d \ln \mu} &=& -g_{um}^2,\nonumber\\
\frac{d g_{um} }{d \ln \mu} &=& -g_{um} g_\rho,\nonumber\\
\label{RG-c}
\end{eqnarray}
in the charge sector and 
\begin{eqnarray}
\frac{d g_\sigma^\perp}{d \ln \mu} &=& -g_\sigma^\perp g_{\sigma}^z,\nonumber\\
\frac{d g_{\sigma}^z}{d \ln \mu} &=& -(g_\sigma^\perp)^2,\nonumber\\
\label{RG-s}
\end{eqnarray} 
in the spin sector.  

\begin{figure}[t]
\begin{center}
%\vspace{0.2cm}
\includegraphics[width=9 cm,clip]{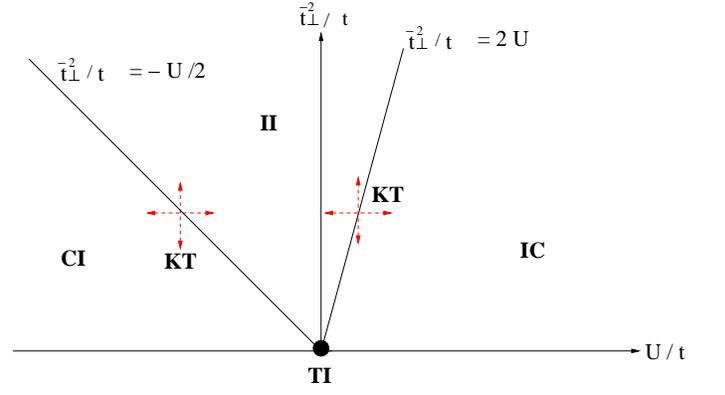}
\end{center}
\par
 \vskip -0.2cm
\caption{
(Color online)
Quantum phase diagram of the zigzag Kane-Mele Hubbard 
ribbon for $N=odd$ at half filling as a function of $U/t$ and 
$\bar{t}_\perp^2/t$. The helical topological edge states (TI) are unstable 
against any $U\neq 0$ or $\bar{t}_\perp\neq 0$, and 
towards the IC, CI and II phases 
for $U>\bar{t}_\perp^2/(2t)$, $U<-2\bar{t}_\perp^2/t$ and 
$-2 \bar{t}_\perp^2/t< U< \frac{1}{2}\bar{t}_\perp^2/t$,  
respectively. The II-IC and II-CI phase transitions are 
of the Kosterlitz-Thouless (KT) type (red dashed arrows).}
\label{phase-half-odd}
\end{figure}

As shown in Figs.~\ref{RG-half-grhogum-odd} and ~\ref{RG-half-gpergz-odd}, 
the generic RG flows of Eqs.~(\ref{RG-c}) and (\ref{RG-s}) are of the 
Kosterlitz-Thouless (KT) type. In the charge sector, the RG flows for 
$g_{um}$ and $g_{\rho}$ with the bare couplings 
$(g_{um}^0, g_{\rho}^0) = (-\bar{t}_\perp^2/t,U/2)$ are always towards either 
the strong-coupling charg and spin gaped II phase for 
$-2 \bar{t}_\perp^2/t< U< \frac{1}{2}\bar{t}_\perp^2/t$ or towards 
the charge conducting and spin insulating CI phase for 
$U< -2 \bar{t}_\perp^2/t$. Similarly, 
in the spin sector, the TI phase is unstable 
against either the II phase for   
$-g_\sigma^{z,0} < g_\sigma^{\perp,0}$ 
({\it ie.} $\bar{t}_\perp^2/t> 2U$) or against a charge gaped but spin gapless  
IC phase for $-g_\sigma^{z,0} > g_\sigma^{\perp,0}$ 
({\it ie.} $\bar{t}_\perp^2/t <2U$) (see Fig.~\ref{RG-half-gpergz-odd}). 
Therefore, the TI phase is unstable against any infinitesmall $U\neq 0$ 
and $\bar{t}_\perp\neq 0$. The II-IC and II-CI quantum phase transitions 
are of the KT type. 
Combining the RG flows for both spin and charge sectors, we obtain 
the global phase diagram shown in Fig. ~\ref{phase-half-odd} 
for $N=odd$ and at half filling:
\begin{figure}[t]
\begin{center}
\includegraphics[width=8cm,clip]{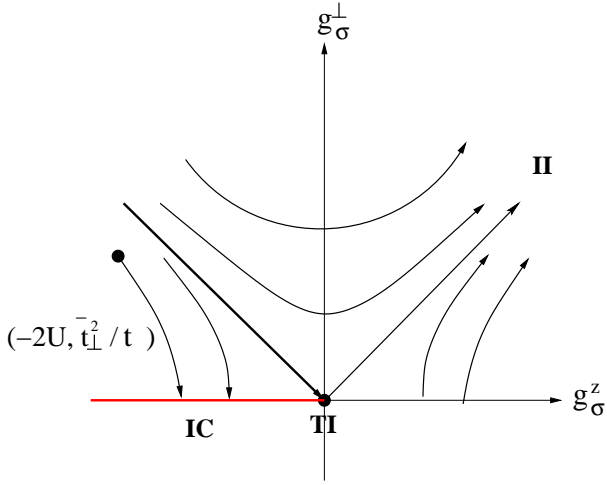}
\end{center}
\par
 \vskip -0.7cm
\caption{
(Color online)
The RG flows of the Kosterlitz-Touless type 
for the spin sector ($g_\sigma^\perp, g_\sigma^z$) 
of the zigzag Kane-Mele Hubbard ribbon for $N=odd$ 
at half filling. The black circle stands for the 
initial (bare) couplings at 
$(g_\sigma^{z,0}, g_\sigma^{\perp,0}) = (-2U, \bar{t}_\perp^2/t)$. 
The arrows indicate the directions of the RG flows upon decreasing 
the curt-off scale $\mu$ from $\mu_0$. 
The red line represents a line of fixed points in the IC phase, 
the II-IC phase boundary is defined by the separatrix line (thick black 
arrow) and its quantum transition is of the Kosterlitz-Thouless (KT) type. 
The topological TI phase is stable only at the origin $U=0=\bar{t}_\perp$.}
\label{RG-half-gpergz-odd}
\end{figure}
\begin{eqnarray}
II&:& -2 \bar{t}_\perp^2/t< U< \frac{1}{2}\bar{t}_\perp^2/t,\nonumber \\
& & g_\rho(\mu\rightarrow 0), g_{um}(\mu\rightarrow 0) \rightarrow \infty,\nonumber \\
 & & \bar{t}_\perp^2/t> 2U, g_\sigma^{z}(\mu\rightarrow 0), g_\sigma^{\perp}(\mu\rightarrow 0) \rightarrow \infty.\nonumber \\
& & \nonumber \\
IC&:& U > \frac{1}{2} \bar{t}_\perp^2/t>0,\nonumber \\
& & g_\sigma^{z}(\mu\rightarrow 0), g_\sigma^{\perp}(\mu\rightarrow 0) \rightarrow 0,\nonumber \\
& & g_\rho(\mu\rightarrow 0), g_{um}(\mu\rightarrow 0) 
\rightarrow \infty.\nonumber \\
& & \nonumber \\
CI&:& U< -2\bar{t}_\perp^2/t <0,  \nonumber \\
& & g_\sigma^{z}(\mu\rightarrow 0), g_\sigma^{\perp}(\mu\rightarrow 0) \rightarrow \infty,\nonumber \\
& & g_\rho(\mu\rightarrow 0), g_{um}(\mu\rightarrow 0) 
\ll 1.\nonumber \\
\label{IC-CI}
\end{eqnarray} 
The topological edge states (TI) are unstable against 
the charge and spin insulating II phase for 
$-2 \bar{t}_\perp^2/t< U< \frac{1}{2}\bar{t}_\perp^2/t$, 
against the charge insulating abd spin conducting IC 
phase for $U > \frac{1}{2} \bar{t}_\perp^2/t>0$, 
and against the charge conducting and spin insulating 
CI phase for $ U< -2\bar{t}_\perp^2/t <0$. 
Therefore, TI phase is unstable for 
any $U\neq 0$ or $\bar{t}_\perp\neq 0$. The II-IC and II-CI quantum 
phase transitions are of the KT type.
Our results on the stability of the TI phase for KM Hubbard model 
on a zigzag ribbon are different from 
those in Ref.~\onlinecite{Teo} through bosonizing the infinite-sized 
helical Luttinger liquid at a fixed interaction strength 
set by the Luttinger parameter 
$K=\sqrt{\frac{1-\frac{U}{2\pi v_F}}{1+ \frac{U}{2\pi v_F}}}$. 
There, they showed that TI is stable for  $1/2<K<2$.
The difference lies in the fact that the 
inter-edge tunneling $t_\perp$ 
arised from the finite-size effect plays an important role here while 
it was absent in Ref.~\onlinecite{Teo}.

%$\frac{1}{2t}\bar{t}_\perp^2<U<-2\bar{t}_\perp^2/t$. 
%The global phase diagram of the KM Hubbard model via our weak-coupling RG 
%analysis is shown in Fig. ~\ref{phase-half-odd}. 
%Our results obtained here based on the weak-coupling RG are 
%qualitatively consistent with those in Ref.~(\onlinecite{Teo}) 
%via bosonization at a finite interaction strength given by the 
%Luttinger parameter 
%$K=\sqrt{\frac{1-\frac{U}{2\pi v_F}}{1+ \frac{U}{2\pi v_F}}}$ though 
%details of the phase boundaries are different: the IC phase 
%is the ground state for strongly repulsive $U$: $K<1/2$, 
%for a strongly attractive $U$: $K>2$ the CI phase is the ground state. 
%Therefore, the helical edge states are stable for $1/2<K<2$. 

\subsubsection{Away from half filling}
\begin{figure}[t]
\begin{center}
\includegraphics[width=7.5cm,clip]{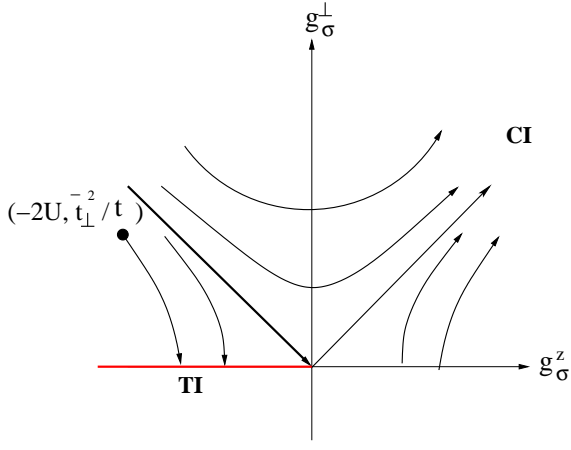}
\end{center}
\par
% \vskip -0.7cm
\caption{
(Color online)
The RG flows of the Kosterlitz-Touless type 
for the spin sector ($g_\sigma^\perp, g_\sigma^z$) 
of the zigzag Kane-Mele Hubbard ribbon away from half filling 
for $N=odd$. 
The black circle stands for the 
initial (bare) couplings. The arrows indicate the directions 
of the RG flows upon decreasing the curt-off scale $\mu$ from $\mu_0$. 
The red line represents a line of fixed points in the TI phase, 
the TI-CI phase boundary is defined by the separatrix line (thick black 
arrow). Note that the coupling $g_\rho$ does not flow under RG 
in this case (see Eq.~(\ref{RG-c-doped})).}
\label{RG-away-gpergz-odd}
\end{figure}
We now proceed to address the case of finite doping away from 
half filling, $k_F a \neq \pi$. 
In this case, the Umklapp term $H_{um}$ vanishes as mentioned in Sec. II. 
The RG scaling equations reduce to:
\begin{eqnarray}
\frac{d g_\rho}{d \ln \mu} &=& 0,\nonumber\\
\label{RG-c-doped}
\end{eqnarray}
in the charge sector with $g_\rho^0 = U$ and 
\begin{eqnarray}
\frac{d g_\sigma^\perp}{d \ln \mu} &=& -g_\sigma^\perp g_{\sigma}^z,\nonumber\\
\frac{d g_{\sigma}^z}{d \ln \mu} &=& -(g_\sigma^\perp)^2,\nonumber\\
\label{RG-s-doped}
\end{eqnarray} 
in the spin sector with $(g_\sigma^{z,0}, g_\sigma^{\perp,0}) = 
(-2U, \bar{t}_\perp^2/t)$. 

\begin{figure}[t]
\begin{center}
%\vspace{0.2cm}
\includegraphics[width=6.5 cm,clip]{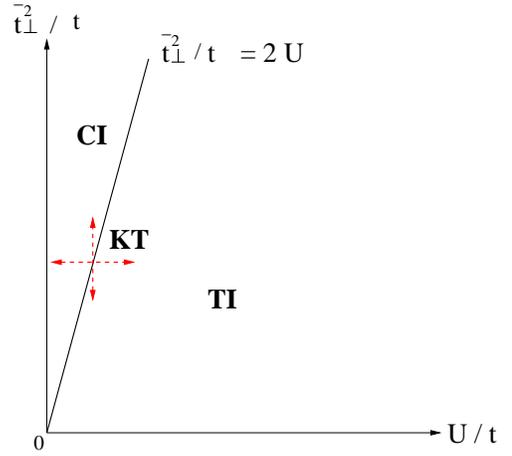}
\end{center}
\par
 \vskip -0.7cm
\caption{
(Color online)
Quantum phase diagram of the zigzag Kane-Mele Hubbard 
ribbon away from half filling for $N=odd$ 
as functions of $\bar{t}_\perp^2/t$ and $U/t$. 
The helical topological edge states (TI phase) are unstable towards 
the charge conducting and spin insulating  
CI phase for $\bar{t}_\perp^2/t>2U$. 
The TI-CI quantum phase transition set by the boundary 
$\bar{t}_\perp^2/t=2U$ 
is of the Kosterlitz-Thouless (KT) type (red dashed arrows).}
\label{phase-away-odd}
\end{figure}

Via Eq.~(\ref{RG-c-doped}), it is clear that  
the system will not develop a charge gap under RG as 
$g_\rho$ does not diverge: $g_\rho(\mu)=g_\rho^0\ll 1$. 
The RG flows in the spin sector (see Eq.~\ref{RG-s-doped}), 
however, suggest that the topological 
edge states may undergo the Kosterlitz-Thouless 
 transition upon increasing $\bar{t}_\perp$ to a spin gaped phase. 
Combing RG flows in both charge and spin sectors, this spin gaped phase 
corresponds to the charge conducting but spin insulating (or CI) phase 
(see Fig. ~\ref{RG-away-gpergz-odd}). 
The TI-CI phase boundary is set by the separatrix  
$g_\sigma^\perp + g_\sigma^z=0$ (or when $\bar{t}_\perp^2/t = 2U$, 
see Fig. ~\ref{RG-away-gpergz-odd}).
The helical edge states are therefore stable 
for $\bar{t}_\perp^2/t < 2U$, while it is unstable against the CI phase for $U< 
\frac{\bar{t}_\perp^2}{2t}$ (see Fig.~\ref{phase-away-odd}). 
\section{Instabilities, orderings, and correlation functions of the Kane-Mele-Hubbard model}

We now investigate further the nature of the TI, CI, IC and II phases. 
In particular, we focus on instabilities towards various orderings 
and correlation 
functions in these phases. 
Various correlation functions with specific orderings can be defined for 
this purpose: (i). the charge-density-wave $\mathcal{O}_{CDW}$ correlation, 
(ii). the spin-density-wave $\mathcal{O}_{SDW}^{a=x,y,z}$ correlation, (iii). 
the singlet $\mathcal{O}_{SS}$ and triplet $\mathcal{O}_{TS}^{a=x,y,z}$ 
superconducting pairing operators, where\cite{giamarchi}
\begin{eqnarray}
\mathcal{O}_{CDW} &=& \Psi^{\dagger\uparrow}_{R,1}(x) \Psi_{L,2}^{\uparrow}(x) + 
\Psi^{\dagger\downarrow}_{R,2}(x) \Psi_{L,1}^{\downarrow}(x),\nonumber \\
\mathcal{O}_{SDW}^{x} &=& \Psi^{\dagger\uparrow}_{R,1}(x) 
\Psi_{L,1}^{\downarrow}(x) + 
\Psi^{\dagger\downarrow}_{R,2}(x) \Psi_{L,2}^{\uparrow}(x),\nonumber \\
\mathcal{O}_{SDW}^{y} &=& -{\it i} 
[\Psi_{R,1}^{\dagger\uparrow}(x) \Psi_{L,1}^{\downarrow}(x) - 
\Psi^{\dagger\downarrow}_{R,2}(x) \Psi_{L,2}^{\uparrow}(x)],\nonumber \\
\mathcal{O}_{SDW}^{z} &=&  
\Psi_{R,1}^{\dagger\uparrow}(x) \Psi_{L,2}^{\uparrow}(x) - 
\Psi^{\dagger\downarrow}_{R,2}(x) \Psi_{L,1}^{\downarrow}(x),\nonumber \\
\mathcal{O}_{SS} &=& \Psi^{\dagger\uparrow}_{R,1}(x) 
\Psi_{L,1}^{\dagger\downarrow}(x) + 
\Psi^{\dagger\uparrow}_{L,2}(x) \Psi_{R,2}^{\dagger\downarrow}(x),\nonumber \\
\mathcal{O}_{TS}^{x} &=& \Psi^{\dagger\uparrow}_{R,1}(x) 
\Psi_{L,2}^{\dagger\uparrow}(x) + 
\Psi^{\dagger\downarrow}_{L,1}(x) \Psi_{R,2}^{\dagger\downarrow}(x),\nonumber \\
\mathcal{O}_{TS}^{y} &=& -{\it i} 
[\Psi_{R,1}^{\dagger\uparrow}(x) \Psi_{L,2}^{\dagger\uparrow}(x) - 
\Psi^{\dagger\downarrow}_{L,1}(x) \Psi_{R,2}^{\dagger\downarrow}(x)],\nonumber \\
\mathcal{O}_{TS}^{z} &=&  
\Psi_{R,1}^{\dagger\uparrow}(x) \Psi_{L,1}^{\dagger\downarrow}(x) - 
\Psi^{\dagger\uparrow}_{L,2}(x) \Psi_{R,2}^{\dagger\downarrow}(x).\nonumber \\
\label{instability-OP}
\end{eqnarray}
Note that some of the operators defined above involve 
helical electrons on both edges, different from those defined for a 
standard Luttinger liquid in one-dimensional interacting electrons 
where all electrons are along the same one-dimensional 
wire\cite{gogolin,giamarchi}. 
To investigate the above correlation functions, it is useful to 
bosonize the Hamiltonian Eq.~(\ref{H_Hubbard}) as\cite{Teo}:
\begin{eqnarray}
H_{edge}^{eff} &=& \int dx [\sum_{\alpha =c,s}\frac{v_{\alpha}}{2} 
 \left(K_{\alpha}(\partial_x\Theta_{\alpha})^2 \! +\frac{1}{K_{\alpha}}
 (\partial_x\Phi_{\alpha})^2\right)\nonumber \\
&+& \frac{t_\perp}{2\pi a_0} \cos(\sqrt{2\pi}\Phi_c +  2k_F x) 
\cos(\sqrt{4\pi} \Phi_s)\nonumber \\
&+&\frac{g_\sigma^\perp}{(2\pi a_0)^2} \cos(2\sqrt{2\pi}\Phi_s) \nonumber \\
&+& \frac{1}{8\pi} 
\frac{g_\sigma^z}{(2\pi a_0)^2} (\partial_x \Phi_s - \partial_x \Theta_s)\nonumber \\
&+& \frac{g_{um}}{(2\pi a_0)^2}  \cos(2\sqrt{2\pi} \Phi_c + 4k_F x)\nonumber \\
&+& 
\frac{1}{4}\frac{g_\rho}{(2\pi a_0)^2}\sum_{\alpha =c,s} 
\left((\partial_x\Phi_{\alpha})^2 - (\partial_x\Theta_{\alpha})^2\right)]\nonumber \\
\label{H_bosonized}
\end{eqnarray}
where via bosonization formulas,\cite{Teo,gogolin,giamarchi}
\begin{eqnarray*}
 \Psi_{L\sigma} &=& \frac{1}{\sqrt{2\pi a_0}}\eta_{\sigma}
 e^{-i\sqrt{4\pi}\phi_{L\sigma}} \ , \\
 \Psi_{R\sigma} &=& \frac{1}{\sqrt{2\pi a_0}}\eta_{\sigma}
 e^{i\sqrt{4\pi}\phi_{R\sigma}} \ ,
\end{eqnarray*}
and the bosonic fields defined as:
\begin{eqnarray*}
 \Phi_{\sigma}&=&\phi_{L\sigma}+\phi_{R\sigma} \ , ~~
 \Theta_{\sigma}=\phi_{L\sigma}-\phi_{R\sigma} \ ,\nonumber \\
\Phi_{c(s)}&=&\frac{1}{\sqrt{2}} (\phi_{\uparrow}\pm \phi_{\downarrow}) \ , ~~
 \Theta_{c(s)}= \frac{1}{\sqrt{2}} 
(\Theta_{\uparrow}\pm \Theta_{\downarrow}).\nonumber \\
\label{Phi-Theta}
\end{eqnarray*}
 with $\eta_\sigma$ being the Klein factor and 
$a_0$ being the short-distance cutoff. In terms of these boson fields,  
the correlation functions mentioned above 
are given by:

\begin{eqnarray}
\mathcal{O}_{CDW} &=& \frac{e^{-2{\it i}k_F x}}{\pi a_0}   
e^{-{\it i}\sqrt{2\pi}\Phi_c} \cos(\sqrt{2\pi} \Phi_s)
,\nonumber \\
\mathcal{O}_{SDW}^{x} &=& \frac{e^{-2{\it i}k_F x}}{\pi a_0}   
e^{-{\it i}\sqrt{2\pi}\Phi_c} \cos(\sqrt{2\pi} \Theta_s),\nonumber \\
\mathcal{O}_{SDW}^{y} &=& -\frac{e^{-2{\it i}k_F x}}{\pi a_0}   
e^{-{\it i}\sqrt{2\pi}\Phi_c} \sin(\sqrt{2\pi} \Theta_s),
\nonumber \\
\mathcal{O}_{SDW}^{z} &=&  {\it{i}}\frac{e^{-2{\it i}k_F x}}{\pi a_0}   
e^{-{\it i}\sqrt{2\pi}\Phi_c} \sin(\sqrt{2\pi} \Phi_s),\nonumber \\
\mathcal{O}_{SS} &=& \frac{1}{\pi a_0}   
e^{{\it i}\sqrt{2\pi}\Theta_c} \cos(\sqrt{2\pi} \Phi_s),\nonumber \\
\mathcal{O}_{TS}^{x} &=& 
\frac{1}{\pi a_0}   
e^{{\it i}\sqrt{2\pi}\Theta_c} \cos(\sqrt{2\pi} \Theta_s),\nonumber \\
\mathcal{O}_{TS}^{y} &=& -\frac{1}{\pi a_0}   
e^{{\it i}\sqrt{2\pi}\Theta_c} \sin(\sqrt{2\pi} \Theta_s),\nonumber \\
\mathcal{O}_{TS}^{z} &=&  \frac{1}{\pi a_0}   
e^{{\it i}\sqrt{2\pi}\Theta_c} \sin(\sqrt{2\pi} \Phi_s).\nonumber \\
\label{instability-OP-HLL}
\end{eqnarray}
Based on the phase diagram via weak-coupling RG and the bosonized 
form of the Hamiltonian, we analyze below the instabilities and the 
behaviors of various correlation functions for (i) the charge and spin 
gapless (TI) topological edge states, (ii) the CI phase, 
(iii) the IC phase, and (iv) the II phase. 

\subsubsection{The topological edge states (TI) phase}
In the gapless topological edge states-- the charge and spin conducting  
state--various correlation functions 
can be computed via correlation functions of the boson fields, given by:
\begin{eqnarray}
<\mathcal{O}_{CDW}^\dagger (0) \mathcal{O}_{CDW}(r) > 
&\sim& e^{-2{\it i}k_F x} 
(\frac{1}{r})^{K_c+K_s} \nonumber \\
&\sim& e^{-2{\it i}k_F x}
(\frac{1}{r})^{1/K +K},\nonumber \\
 <\mathcal{O}_{SDW}^{\dagger x} (0) \mathcal{O}_{SDW}^x(r) > 
&\sim& e^{-2{\it i}k_F x}
(\frac{1}{r})^{K_c+1/K_s} \nonumber \\
&\sim& e^{-2{\it i}k_F x}
(\frac{1}{r})^{2K},\nonumber \\
 <\mathcal{O}_{SDW}^{\dagger y} (0) \mathcal{O}_{SDW}^y(r) > 
&\sim& e^{-2{\it i}k_F x}
(\frac{1}{r})^{K_c+1/K_s} \nonumber \\
&\sim& e^{-2{\it i}k_F x}
(\frac{1}{r})^{2K},\nonumber \\
 <\mathcal{O}_{SDW}^{\dagger z} (0) \mathcal{O}_{SDW}^z(r) > 
&\sim& e^{-2{\it i}k_F x}
(\frac{1}{r})^{K_c+K_s} \nonumber \\
&\sim& e^{-2{\it i}k_F x}
(\frac{1}{r})^{1/K+K},\nonumber \\
<\mathcal{O}_{SS}^\dagger (0) \mathcal{O}_{SS}(r) > 
&\sim& (\frac{1}{r})^{1/K_c+K_s} \sim (\frac{1}{r})^{2/K},\nonumber \\
 <\mathcal{O}_{TS}^{\dagger x} (0) \mathcal{O}_{TS}^x(r) > 
&\sim& (\frac{1}{r})^{1/K_c+1/K_s} \sim (\frac{1}{r})^{K+1/K},\nonumber \\
 <\mathcal{O}_{TS}^{\dagger y} (0) \mathcal{O}_{TS}^y(r) > 
&\sim& (\frac{1}{r})^{1/K_c+1/K_s} \sim (\frac{1}{r})^{1/K+K},\nonumber \\
 <\mathcal{O}_{TS}^{\dagger z} (0) \mathcal{O}_{TS}^z(r) > 
&\sim& (\frac{1}{r})^{1/K_c+K_s} \sim (\frac{1}{r})^{1/(2K)}\nonumber \\
\label{correlation-TI}
\end{eqnarray}
with $K_c=K$ and $K_s=1/K$ in the helical Luttinger liquid\cite{Teo}. 
Note that in the conventional spinful Luttinger liquids where $K_s=1$, 
the above correlation functions get modified accordingly\cite{bena,gogolin,giamarchi}.

\subsubsection{The CI phase}
Now, we analyze instability and correlation functions 
in the the charge conducting and spin insulating 
(CI) phase. As shown in Eqs.~(\ref{CI}) and (\ref{H_bosonized}), 
$g_\sigma^\perp, g_\sigma^z \rightarrow \infty$ while 
$g_\rho,g_{um}\rightarrow 0$  
in this phase. In the bosonized form of the Hamiltonian, this implies 
that $\Phi_s$ is pinned to a constant value\cite{Teo,gogolin,giamarchi}: 
$\Phi_s\sim n \pi/\sqrt{8\pi}$. 
As a result, its conjugate variable $\Theta_s$ is disordered and 
exhibit exponentially decaying correlation functions\cite{gogolin,giamarchi}. 
The corresponding leading correlation functions have the following power-law 
behaviors:
\begin{eqnarray}
<\mathcal{O}_{CDW}^\dagger (0) \mathcal{O}_{CDW}(r) > 
&\sim& (\frac{1}{r})^{K_c} \sim (\frac{1}{r})^{K},\nonumber \\
<\mathcal{O}_{SS}^\dagger (0) \mathcal{O}_{SS}(r) > 
&\sim& (\frac{1}{r})^{1/K_c} \sim (\frac{1}{r})^{1/K}.\nonumber \\
% <\mathcal{O}_{TS}^{\dagger z} (0) \mathcal{O}_{TS}^z(r) >. \nonumber \\
%&\sim& (\frac{1}{r})^{1/K_c} \sim (\frac{1}{r})^{1/K}.\nonumber \\
\label{correlation-TI}
\end{eqnarray}
Note that due to the disordered nature of the $\Theta_s$ field, 
the SDW as well as the  
TS orderings vanish: 
$<\mathcal{O}_{SDW}^{\dagger x,y,z}\mathcal{O}_{SDW}^{x,y,z}>\rightarrow 0, 
<\mathcal{O}_{TS}^{\dagger x,y,z} \mathcal{O}_{TS}^{x,y,z}> \rightarrow 0$.
Therefore, we find the leading instabilities of the CI phase are towards 
the CDW and superconductivity (SC). For repulsive interactions $K<1$ (or $U>0$) 
that we consider here, the CDW order is dominating over the SC order 
as CDW correlators decay more slowly than that for SC orders. However, for attractive interactions $K>1$ (or $U<0$), it is the SC order which 
dominates the CI phase.

\subsubsection{The IC phase}
We now analyze the instability of the charge insulating but 
spin conducting (IC) phase. It is clear from 
Eq.~(\ref{H_bosonized}) that $\Phi_c$ field is pinned to a 
constant value in this phase: $\Phi_c \sim n\pi/\sqrt{8\pi}$. 
The correlation functions for the CDW and SDW orderings are given by:
\begin{eqnarray}
<\mathcal{O}_{CDW}^\dagger (0) \mathcal{O}_{CDW}(r) > 
&\sim& (\frac{1}{r})^{K_s} \sim (\frac{1}{r})^{1/K},\nonumber \\
 <\mathcal{O}_{SDW}^{\dagger x} (0) \mathcal{O}_{SDW}^x(r) > 
&\sim& (\frac{1}{r})^{1/K_s} \sim (\frac{1}{r})^{K},\nonumber \\
 <\mathcal{O}_{SDW}^{\dagger y} (0) \mathcal{O}_{SDW}^y(r) > 
&\sim& (\frac{1}{r})^{1/K_s} \sim (\frac{1}{r})^{K},\nonumber \\
 <\mathcal{O}_{SDW}^{\dagger z} (0) \mathcal{O}_{SDW}^z(r) > 
&\sim& (\frac{1}{r})^{K_s} \sim (\frac{1}{r})^{1/K}.\nonumber \\
\label{correlation-TI}
\end{eqnarray}
On the other hand, due to the 
pinning of the $\Phi_c$ field, its conjugate 
field $\Theta_c$ is completely disordered. Hence, the SS and TS 
orderings are suppressed: $<\mathcal{O}_{SS}^{\dagger} 
\mathcal{O}_{SS}> \rightarrow 0$, $<\mathcal{O}_{TS}^{\dagger x,y,z} 
\mathcal{O}_{TS}^{x,y,z}> \rightarrow 0$. For repulsive Hubbard term $U>0$ 
(or $K<1$), the SDW orderings along 
$x-$ and $y-$ directions are the leading instabilities of this phase 
as their correlation functions decay more slowly compared to the 
others. The system shows quasi-long-ranged magnetic order. 
This phase shares similarities to the Mott insulating 
phase in the sense that interactions lead to a metal-insulator 
transition and at the same time to a state with 
magnetic order. In fact, this phase corresponds 
to the SDW phase found in the mean-field approach of the KM Hubbard 
in Ref.~\onlinecite{hur}. 
For the 
attractive Hubbard model $U<0$ (or $K>1$), however, 
the leading instabilities go towards the 
CDW and SDW along the $z-$axis . 

\subsubsection{The II phase}

Finally, we analyze the charge and spin insulating II phase. 
This phase occurs for a finite-sized ribbon at half filling 
where all the couplings--the inter-edge hopping term $t_\perp$, 
the Umklapp term $g_{um}$, scalar density-density interaction $g_\rho$, 
the two-particle spin scattering terms $g_\sigma^{\perp,z}$-- 
become relevant under RG, 
$t_\perp,g_{um,\rho},g_\sigma^{\perp,z} \rightarrow \infty$. 
From the bosonized Hamiltonian 
Eq.~(\ref{H_bosonized}), this phase requires the pinning of 
both $\Phi_c$ and $\Phi_s$ fields at $\Phi_{c,s}\approx n\pi/\sqrt{2\pi}$, 
leading to exponential decay 
of all the correlation functions associated with the orderings 
in Eq.~(\ref{instability-OP}) except for the CDW ordering with a 
constant correlator. Whether or not 
the II phase found here is related to  
the spin-gaped, charge-gaped (similar to II phase)  
spin-liquid phase found numerically via QMC in Ref.~\onlinecite{Meng} 
or furthermore to the Anderson's resonant-valence-bond (RVB) spin liquid  
need further investigations.

\section{Conclusions.} 

In summary, we have studied the stability of the helical edge states 
and quantum phases and phase transitions 
of the Kane-Mele Hubbard (KMH) model on a finite-sized zigzag ribbon 
of honeycomb lattice. 

We first focus on the finite-size effect of the Kane-Mele (KM) 
zigzag ribbon in the absence of the on-site Hubbard interaction.  
We first reproduce in the energy excitation spectrum 
the well-known Dirac-dispersed topological edge states. 
 In additions, due to the finite ribbon size, we 
have shown that a finite inter-edge hopping between two 
edge states exist, which falls off exponentially with increasing 
ribbon width. This inter-edge hopping term 
generates via second order perturbation 
two important two-particle scatterings: the inter-edge spin-flip 
term and the inter-edge backscattering (or the Umklapp term). 
These three terms lead to instabilities of the topological 
edge states. 
 
We further analyze the instabilities of the topological edge states, 
as well as possible quantum phases and phase transitions 
upon including a weak on-site repulsive 
Hubbard interaction on the zigzag KM ribbon. 
Via perturbative RG approach we find 
the combined effects from 
the inter-edge hopping and the on-site Coulomb interactions lead to the 
instabilities of the topological edge states (TI phase) 
against (i) the charge and spin insulating II phase, 
(ii) the charge insulating but spin conducting IC phase, and (iii)
the charge conducting but spin insulating CI phase, 
depending on $N=even/odd$, the electron density (filling factor), 
and on the ratio of 
the Coulomb interaction $U$ and the inter-edge tunneling $t_\perp$, 
$U/t_\perp$. Via RG analysis we find the quantum 
phase transitions for TI-CI, II-IC and II-CI are of the 
Kosterlitz-Thouless type. Via bosonization approach, we furthermore 
investigated the instabilities towards new orderings, including the 
CDW, SDW and superconducting orders by computing 
correlation functions of these orderings in the helical 
edge states, as well as in the CI, IC, and II phases. Our theoretical 
predictions can serve as a basis to investigate further both theoretically 
and experimentally correlation effects or Mott physics 
in interacting topological insulators.

%%%%%%%%%%%%%%%%%%%%%%%%%%%%%%%%%%%%%%%%%%%%%%%%%%%%%%%%%%%%%%%%%%%%%%%

%\subsection{ IV. Conclusions.}

\acknowledgements
We acknowledge M. Cazalilla, C.Y. Mou   
for helpful discussions. 
This work is supported by the NSC grant
No.98-2918-I-009-06, No.98-2112-M-009-010-MY3, the NCTU-CTS, 
the MOE-ATU program, the NCTS of Taiwan, R.O.C..

%%%%%%%%%%%%%%%%%%%%%%%%%%%%%%%%%%%%%%%%%%%%%%%%%%%%%%%%%%%%%%%%%%%%%%%

%\appendix 

%\section{The RG scaling equation in the weak-coupling regime}

\end{document}